\RequirePackage[l2tabu]{nag}		
\documentclass[a4paper,11pt,leqno,openbib,oldfontcommands]{memoir} 
\usepackage[inkscapelatex=false]{svg}

\usepackage{pgfplots}
\usepackage{tikz}
\pgfplotsset{compat=1.18}

\usepackage{amsmath,amssymb,amsfonts}
\usepackage{algorithmic}
\usepackage{graphicx}
\usepackage{textcomp}
\usepackage{xcolor, soul}
\sethlcolor{red}
  \def\BibTeX{{\rm B\kern-.05em{\sc i\kern-.025em b}\kern-.08em
    T\kern-.1667em\lower.7ex\hbox{E}\kern-.125emX}}

\usepackage{listings}
\usepackage{color}

\definecolor{codegreen}{rgb}{0,0.6,0}
\definecolor{codegray}{rgb}{0.5,0.5,0.5}
\definecolor{codepurple}{rgb}{0.58,0,0.82}
\definecolor{backcolour}{rgb}{0.95,0.95,0.92}

\lstdefinestyle{mystyle}{
    backgroundcolor=\color{backcolour},   
    commentstyle=\color{codegreen},
    keywordstyle=\color{magenta},
    numberstyle=\tiny\color{codegray},
    stringstyle=\color{codepurple},
    basicstyle=\ttfamily\footnotesize,
    breakatwhitespace=false,         
    breaklines=true,                 
    captionpos=b,                    
    keepspaces=true,                 
    numbers=left,                    
    numbersep=5pt,                  
    showspaces=false,                
    showstringspaces=false,
    showtabs=false,                  
    tabsize=2
}

\usepackage{datetime}
\usepackage[ruled,vlined]{algorithm2e}
\usepackage{ifpdf}
\ifpdf
\fi
\ifdraftdoc 
	\usepackage{draftwatermark}				
	\SetWatermarkScale{0.3}
	\SetWatermarkText{\bf Draft: \today}
\fi
\settrimmedsize{297mm}{210mm}{*}
\settypeblocksize{634pt}{448.13pt}{*} 
\setulmargins{3cm}{*}{*} 
\setlrmargins{*}{*}{1.5} 
\setmarginnotes{17pt}{51pt}{\onelineskip} 
\setheadfoot{\onelineskip}{2\onelineskip} 
\setheaderspaces{*}{2\onelineskip}{*} 
\checkandfixthelayout
\OnehalfSpacing 
\setsecnumdepth{subsection} 
\maxsecnumdepth{subsubsection}
\makepagestyle{myvf} 
\makeoddfoot{myvf}{}{\thepage}{} 
\makeevenfoot{myvf}{}{\thepage}{} 
\makeheadrule{myvf}{\textwidth}{\normalrulethickness} 
\makeevenhead{myvf}{\small\textsc{\leftmark}}{}{} 
\makeoddhead{myvf}{}{}{\small\textsc{\rightmark}}
\newcommand{\clearemptydoublepage}{\newpage{\thispagestyle{empty}\cleardoublepage}}
\makeindex
\usepackage{import}
\usepackage{multicol}
\usepackage{geometry}
\usepackage{amsfonts} 					
\usepackage{amsmath}			
\usepackage{stmaryrd}					
\usepackage{amssymb}					
\usepackage{amsthm}					
\usepackage{newlfont}					
\usepackage{layouts}					
\usepackage{graphicx}					
\usepackage[utf8]{inputenc}			
\usepackage{float}						
\usepackage[square,numbers,
		     sort&compress]{natbib}		
\usepackage{url}						
\usepackage[spanish,english]{babel}		

\usepackage[colorlinks=true,
		     allcolors=black]{hyperref}              
\usepackage{memhfixc}					
\usepackage{enumerate}					
\usepackage{footnote}					
\usepackage{microtype}					

\usepackage{listings}
\usepackage{setspace}

\usepackage{subcaption}
\usepackage{xcolor}
\definecolor{UniversityRed}{RGB}{171,31,45}
\theoremstyle{plain}

\theoremstyle{plain}

\theoremstyle{plain}
\theoremstyle{definition}

\theoremstyle{plain}

\theoremstyle{plain}

\theoremstyle{plain}

\begin{document}
%
%
%
%
%
\frontmatter
\pagenumbering{roman}
\newgeometry{left=2cm, right=2cm, top=4cm, bottom=3.5cm}
\begin{multicols}{2}
\setlength{\parindent}{0pt}
\footnotesize\textsc{Egypt-Japan University of Science and Technology}\\
\footnotesize\textsc{Faculty of Engineering and Applied Science}\\
\footnotesize\textsc{Mechatronics and Robotics Department}\\

\includegraphics[width = 0.48 \textwidth]{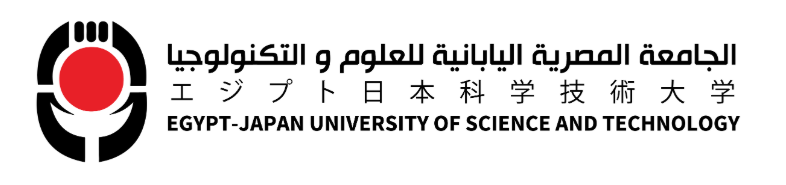}
\end{multicols}
\vspace{1mm}
\vspace{1mm}

\begin{titlingpage}
\begin{SingleSpace}
\calccentering{\unitlength} 
\begin{adjustwidth*}{\unitlength}{-\unitlength}
\vspace*{5mm}
\begin{center}
\rule[0.5ex]{\linewidth}{2pt}\vspace*{-\baselineskip}\vspace*{3.2pt}
\rule[0.5ex]{\linewidth}{1pt}\\[\baselineskip]
{\HUGE Enhancing AUTOSAR-Based Firmware Over-the-Air Updates in the Automotive Industry with a Practical Implementation on a Steering System}\\[8mm]
\rule[0.5ex]{\linewidth}{1pt}\vspace*{-\baselineskip}\vspace{3.2pt}
\rule[0.5ex]{\linewidth}{2pt}\\
\vspace{8mm}
\begin{minipage}{11cm}
\centering
\Large \textsc {A dissertation submitted under the requirements of a Bachelor's Degree in Mechatronics Engineering.}
\end{minipage}\\
\vspace{15mm}
{\Large \textbf{Submitted By}}\\
\vspace{5mm}
\renewcommand{\arraystretch}{1.5}
\begin{tabular}{lclc}
\Large\textsc{\textbf{Mostafa Ahmed Mostafa Ahmed}} & & & \Large\textsc{120200214} \\
\Large\textsc{\textbf{Mohamed Khaled Mohamed Elsayed}} & & & \Large\textsc{120210346} \\
\Large\textsc{\textbf{Radwa Waheed Ezzat Abdelmohsen}} & & & \Large\textsc{120200228} \\
\end{tabular}
\renewcommand{\arraystretch}{1} 
\vspace{15mm}

{\Large \textbf{Supervised By}}\\
\vspace{5mm}
{\Large \textbf{Dr. Mohamed G. Alkalla}}\\
\vspace{3mm}
\begin{minipage}{0.9\textwidth}
\centering
{\small Associate Professor at Mechatronics and Robotics Engineering Department, \\ 
School of Innovative Design Engineering, Egypt-Japan University of Science and Technology, Egypt.}
\end{minipage}
\vspace{10mm}
\vspace{8mm}

{\large \textbf{Academic Year 2024-2025}}

\end{center}
\end{adjustwidth*}
\end{SingleSpace}
\end{titlingpage}
\restoregeometry

\clearemptydoublepage
%
%
%
%

\chapter*{Abstract}
\begin{SingleSpace}

The automotive industry is increasingly reliant on software to manage complex vehicle functionalities, making efficient and secure firmware updates essential. Traditional firmware update methods, requiring physical connections through On-Board Diagnostics (OBD) ports, are inconvenient, costly, and time-consuming. Firmware Over-the-Air (FOTA) technology offers a revolutionary solution by enabling wireless updates, reducing operational costs, and enhancing the user experience. This project aims to design and implement an advanced FOTA system tailored for modern vehicles, incorporating the AUTOSAR architecture for scalability and standardization, and utilizing delta updating to minimize firmware update sizes, thereby improving bandwidth efficiency and reducing flashing times. To ensure security, the system integrates the UDS 0x27 protocol for authentication and data integrity during the update process. Communication between Electronic Control Units (ECUs) is achieved using the CAN protocol, while the ESP8266 module and the master ECU communicate via SPI for data transfer. The system’s architecture includes key components such as a bootloader, boot manager, and bootloader updater to facilitate seamless firmware updates. The functionality of the system is demonstrated through two applications: a blinking LED and a Lane Keeping Assist (LKA) system, showcasing its versatility in handling critical automotive features. This project represents a significant step forward in automotive technology, offering a user-centric, efficient, and secure solution for automotive firmware management.

\BlankLine
\BlankLine
\BlankLine
\BlankLine
\BlankLine
\BlankLine
See \textcolor{blue}{\underline{{\href{https://github.com/Gaafoor/FOTA-and-Remote-Diagnostics}{\textcolor{blue}{https://github.com/Gaafoor/FOTA-and-Remote-Diagnostics}}}}} for code.
\end{SingleSpace}
\clearpage

\clearemptydoublepage
%
%
%

\chapter*{Dedication and Acknowledgements}
\begin{SingleSpace}
First of all, we would like to thank Allah for every progress we have achieved during our respected journey. We would also like to express our heartfelt gratitude to all those who contributed to the completion of this thesis. We extend our deepest appreciation to our supervisor, Dr. Mohamed G. Alkalla, for his invaluable guidance and support, and to the faculty and staff of Egypt-Japan University of Science and Technology for providing the resources and a conducive academic environment. Special thanks to Eng. Ahmed Attia from Valeo for his expert guidance and industry insights, which greatly enriched our work. We are also profoundly thankful to our families and friends for their unwavering encouragement, to our classmates and colleagues for their collaboration and knowledge sharing, and to all who played a role in this journey. Your support has been instrumental in the success of this thesis.

\end{SingleSpace}
\clearpage
\clearemptydoublepage
%
%
%
%
%
%
%
\chapter*{Authors' declaration}
\begin{SingleSpace}
\begin{quote}
I declare that the work in this dissertation was carried out in accordance with the requirements of the University's Regulations and Code of Practice for Bachelor's Degree Programs and that it has not been submitted for any other academic award. Except where indicated by specific reference in the text, the work is the candidate's own work. Work done in collaboration with, or with the assistance of, others, is indicated as such. Any views expressed in the dissertation are those of the author.

\vspace{1.5cm}
\noindent
\hspace{-0.75cm}\textsc{SIGNED: .................................................... DATE: ..........................................}

\noindent
\hspace{-0.75cm}\textsc{SIGNED: .................................................... DATE: ..........................................}

\noindent
\hspace{-0.75cm}\textsc{SIGNED: .................................................... DATE: ..........................................}

\noindent
\hspace{-0.75cm}\textsc{SIGNED: .................................................... DATE: ..........................................}

\noindent
\hspace{-0.75cm}\textsc{SIGNED: .................................................... DATE: ..........................................}

\end{quote}
\end{SingleSpace}
\clearpage
\clearemptydoublepage

\renewcommand{\contentsname}{Table of Contents}
\setcounter{tocdepth}{2} 
\tableofcontents*
\addtocontents{toc}{\par\nobreak \mbox{}\hfill{\bf Page}\par\nobreak}
\clearemptydoublepage
\listoftables
\addtocontents{lot}{\par\nobreak\textbf{{\scshape Table} \hfill Page}\par\nobreak}

\clearemptydoublepage
\listoffigures
\addtocontents{lof}{\par\nobreak\textbf{{\scshape Figure} \hfill Page}\par\nobreak}
\clearemptydoublepage

\chapter*{List of Acronyms}
\addcontentsline{toc}{chapter}{List of Acronyms}

\begin{tabbing}
    \textbf{FOTA} \hspace{2cm} \= Firmware Over-the-Air \\
    \textbf{UDS} \> Unified Diagnostic Services \\
    \textbf{LKA} \> Lane Keeping Assist \\
    \textbf{ECU} \> Electronic Control Unit \\
    \textbf{PID} \> Proportional-Integral-Derivative \\
    \textbf{ADAS} \> Advanced Driver Assistance Systems \\
    \textbf{AUTOSAR} \> Automotive Open System Architecture \\
    \textbf{CAN} \> Controller Area Network \\
    \textbf{SPI} \> Serial Peripheral Interface \\
    \textbf{UART} \> Universal Asynchronous Receiver-Transmitter \\
    \textbf{OBD} \> On-Board Diagnostics \\
    \textbf{RTOS} \> Real-Time Operating System \\
    \textbf{FreeRTOS} \> Free Real-Time Operating System \\
    \textbf{MCU} \> Microcontroller Unit \\
    \textbf{ESP8266} \> Wi-Fi Microcontroller Module \\
    \textbf{STM32} \> STMicroelectronics 32-bit Microcontroller \\
    \textbf{RAM} \> Random Access Memory \\
    \textbf{ROM} \> Read-Only Memory \\
    \textbf{OTA} \> Over-the-Air \\
    \textbf{EEPROM} \> Electrically Erasable Programmable Read-Only Memory \\
\end{tabbing}

%
%
\mainmatter
\chapter{Introduction}
\label{chap: introduction}

\section{Firmware Over-the-Air (FOTA): Revolutionizing Automotive Updates}
\label{sec:fota_intro}
{
In a world where technological advancements shape the automotive industry, keeping vehicles secure, efficient, and up-to-date has become a crucial necessity. Traditional firmware update methods require users to visit repair shops, where vehicles are manually connected to diagnostic equipment via the On-Board Diagnostics (OBD) port. This process is not only inconvenient and costly but also time-intensive, creating a significant barrier to timely updates. Firmware Over-the-Air (FOTA) technology emerges as a groundbreaking solution, addressing these challenges by enabling wireless updates that save time, reduce costs, and enhance user experience.

The primary objective of this project is to design and implement an advanced FOTA system tailored to the evolving demands of the automotive sector. By leveraging the AUTOSAR architecture, the system ensures modularity, scalability, and compliance with industry standards. A key feature of this project is delta updating, which significantly minimizes update size by transmitting only the differences between firmware versions. This optimization not only reduces bandwidth usage but also shortens flashing times, allowing updates to be completed swiftly. Additionally, robust security measures are integrated through the UDS 0x27 protocol, which authenticates the update process and prevents unauthorized access.

The system’s architecture includes essential components such as a bootloader, boot manager, and bootloader updater, which work together to ensure seamless transitions between firmware states and uphold system reliability. Communication between the master and target ECUs is achieved using the CAN protocol, renowned for its robustness and real-time capabilities, while SPI facilitates efficient data transfer between the ESP8266 module and the master ECU. To demonstrate the system's capabilities, two applications have been implemented: a blinking LED application and a lane-keeping assist system, showcasing the versatility and adaptability of the proposed solution.

This research represents a significant leap forward in automotive technology, addressing user-centric challenges while adhering to industry standards. By integrating advanced features and ensuring a seamless user experience, the project highlights the transformative potential of FOTA in modern vehicles.
}
\section{Problem Statement}
\label{sec:PS}
The traditional approach to updating vehicle firmware requires users to visit repair shops to physically connect their vehicles to diagnostic equipment via the On-Board Diagnostics (OBD) port. This process is not only inconvenient but also incurs significant time and monetary costs. As software updates become larger and more frequent, these inefficiencies are further magnified, creating a frustrating user experience and delaying critical updates that ensure vehicle safety and functionality.

Moreover, conventional Firmware Over-the-Air (FOTA) systems often suffer from extended flashing times, further impacting user satisfaction. The reliance on physical connections and lengthy update durations underscore the need for an efficient, secure, and user-centric solution. The growing reliance on complex software within modern vehicles necessitates advancements in firmware update methodologies to maintain optimal performance and security.

This project proposes a robust FOTA system to address these challenges. By incorporating delta updating, the size of firmware updates is significantly reduced, directly minimizing flashing time and bandwidth requirements. Leveraging the AUTOSAR architecture ensures the system's scalability and standardization across diverse vehicle models and ECUs. Security is enhanced through the integration of the UDS 0x27 protocol (Key and Seed), ensuring data integrity and preventing unauthorized access. Reliable communication is achieved through the CAN protocol for ECU interactions and SPI for data transfer between the ESP8266 module and the master ECU. Collectively, these innovations aim to revolutionize the firmware update process, offering a seamless, efficient, and secure solution for modern vehicles.

\section{Research Statement}
\label{sec:RS}
This research aims to develop a comprehensive FOTA solution tailored to the automotive industry. By leveraging the AUTOSAR architecture, the project seeks to establish a robust memory management system capable of handling firmware updates efficiently. The integration of delta updating reduces bandwidth consumption and flashing time, directly addressing user experience concerns. Security is enhanced using the UDS 0x27 security access service, ensuring authentication and data integrity throughout the update process. The system architecture includes a bootloader to manage erasing and writing processes, a boot manager for decision-making, and a bootloader updater to ensure the bootloader remains up-to-date. Communication between ECUs is facilitated through the CAN protocol, ensuring real-time performance and fault tolerance. Additionally, the system incorporates a user-friendly graphical interface for firmware uploads, enabling seamless interaction with the ESP8266 module, which communicates with the master ECU via SPI. The project demonstrates the functionality of the FOTA system through two applications: a simple blinking LED and a lane-keeping assist system, highlighting the adaptability of the proposed solution.

\section{Research Question}
\label{sec:RQ}

\begin{enumerate}

\item \textbf{General Research Questions:}
How can the integration of FOTA enhance vehicle firmware update processes? In what ways does the AUTOSAR architecture contribute to the scalability and modularity of the FOTA system? How does delta updating improve user experience and reduce flashing time during firmware updates?

\item \textbf{Security Mechanisms:}
How effective is the UDS 0x27 protocol (Key and Seed) in ensuring the security and integrity of firmware updates? What measures can be implemented to further enhance the authentication process?

\item \textbf{Communication Protocols:}
How does the CAN protocol ensure reliable and real-time communication between the master and target ECUs? How does SPI communication optimize data transfer between the ESP8266 module and the master ECU?

\item \textbf{System Components:}
How do the bootloader, boot manager, and bootloader updater contribute to the reliability and efficiency of the FOTA process? What are the challenges in implementing these components, and how can they be addressed?

\item \textbf{Applications:}
How effective is the FOTA system in deploying and managing diverse applications, such as the blinking LED and lane-keeping assist system? What are the limitations of the system in handling complex automotive functionalities?

\item \textbf{User Experience and Performance:}
How does the integration of a graphical user interface enhance the usability of the FOTA system? To what extent does the reduced update size and flashing time improve the overall user experience?
\end{enumerate}

\section{Research Objectives}
\label{sec:RO}

\begin{enumerate}
\item To design and implement an AUTOSAR-compliant memory management system for FOTA.
\item To integrate delta updating to minimize the size of firmware updates and optimize data transmission.
\item To enhance system security using the UDS 0x27 protocol (Key and Seed) for authentication and data integrity.
\item To develop a bootloader, boot manager, and bootloader updater to ensure seamless firmware updates and system reliability.
\item To establish reliable communication between the master and target ECUs using the CAN protocol.
\item To enable efficient data transfer between the ESP8266 module and the master ECU using SPI communication.
\item To create a user-friendly graphical interface for firmware uploads, simplifying the update process.
\item To demonstrate the functionality of the FOTA system through two applications: a blinking LED and a lane-keeping assist system.
\end{enumerate}

\section{Research Overview}
\label{sec:ROv}

Enhancing automotive firmware update efficiency and user experience through the integration of Firmware Over-the-Air (FOTA) technology with delta updating, AUTOSAR architecture, and robust communication protocols.
\begin{enumerate}
\item \textbf{Introduction:}
The landscape of automotive technology is rapidly evolving, with increasing reliance on software to control vehicle functionalities. However, traditional firmware update processes often pose significant challenges for users and manufacturers. This research endeavors to explore a transformative solution through Firmware Over-the-Air (FOTA) technology, integrating cutting-edge innovations to enhance user experience and vehicle performance.

\item \textbf{Context and Significance:}
The automotive industry has faced persistent challenges in ensuring timely and efficient firmware updates. Traditional methods, reliant on physical connections through OBD ports, are costly, time-consuming, and inconvenient for users. The proposed FOTA system addresses these issues by leveraging AUTOSAR for standardization and modularity, integrating delta updating to minimize update sizes, and employing robust communication protocols such as CAN and SPI. These innovations collectively enhance user satisfaction, reduce costs, and ensure vehicles remain secure and functional, marking a significant advancement in automotive technology.

\item \textbf{Current Challenges and Technological Gaps:}
Existing FOTA systems often suffer from inefficiencies such as extended flashing times and security vulnerabilities. This research identifies these gaps and proposes solutions through advanced memory management, secure protocols like UDS 0x27, and reliable communication frameworks, paving the way for a seamless update process.

\item \textbf{Methodology:}
The research methodology involves designing and implementing an AUTOSAR-compliant FOTA system. Key components include delta updating for optimized data transmission, a bootloader, boot manager, and bootloader updater for system reliability, and secure communication using CAN and SPI protocols. A graphical user interface further enhances usability, ensuring a user-friendly update process.

\item \textbf{Expected Contributions:}
This research contributes to advancing automotive firmware update processes by demonstrating the efficacy of delta updating, robust communication protocols, and secure authentication mechanisms. The integration of AUTOSAR architecture ensures scalability and standardization, making the system adaptable to various vehicle models. The successful implementation of applications such as a blinking LED and lane-keeping assist highlights the system's versatility.

\item \textbf{Conclusion:}
By addressing the limitations of traditional firmware update methods, this research advances the state of FOTA technology, providing a secure, efficient, and user-friendly solution. The findings are expected to set a new benchmark for automotive firmware management, enhancing both user experience and vehicle functionality.
\end{enumerate}

\clearemptydoublepage

\chapter{Literature Review}
\label{chap: LR}

\section{Related Work}
\label{sec:RW}

{The automotive industry is undergoing a significant transformation driven by advancements in software and connectivity. Firmware Over-the-Air (FOTA) systems have emerged as an essential innovation, enabling manufacturers to remotely update vehicle firmware and enhance vehicle functionality, security, and user satisfaction. This review examines the core concepts and related work that form the foundation of our research, focusing on the integration of FOTA within the AUTOSAR architecture and its implications for memory management, security protocols, and communication frameworks.
}

\subsection{Overview of Firmware Over-the-Air (FOTA) Systems}

Firmware Over-the-Air (FOTA) technology has revolutionized the automotive industry by streamlining the process of delivering firmware updates to vehicles. Unlike traditional methods requiring physical connections through the On-Board Diagnostics (OBD) port, FOTA facilitates wireless updates, reducing time, cost, and inconvenience associated with manual update processes \cite{FOTA_Benefits}. This technology is instrumental in ensuring vehicles remain secure, functional, and updated in a rapidly evolving software-driven environment \cite{AUTOSAR_FOTA}.

A standard FOTA system consists of several key components:

\begin{enumerate} \item \textbf{FOTA Server:} This manages the distribution of firmware updates and securely communicates with vehicles over internet protocols. \item \textbf{FOTA Gateway:} Positioned within the vehicle, the gateway facilitates secure communication between the server and the target Electronic Control Units (ECUs). \item \textbf{Electronic Control Units (ECUs):} These are the end modules that receive, verify, and apply the firmware updates to ensure functionality \cite{ECU_Implementation}. \end{enumerate}

The AUTOSAR framework has further augmented the efficiency of FOTA systems by introducing standardized approaches to managing software updates. This framework enhances modularity and scalability, enabling vehicle manufacturers to deploy FOTA systems across diverse vehicle models while maintaining compliance with industry standards \cite{AUTOSAR_SWS}.

FOTA systems provide numerous advantages, including rapid deployment of software patches, feature enhancements, and security updates without requiring vehicle downtime. Furthermore, the incorporation of delta updates, which transmit only the differences between firmware versions, significantly reduces data transmission costs and update times, making FOTA a cost-effective and convenient solution for manufacturers and users alike \cite{Delta_Update_Advantages}.

\subsection{AUTOSAR Architecture and Memory Management}

{AUTOSAR (AUTomotive Open System ARchitecture) has become a cornerstone in the evolution of automotive software systems, providing a modular and scalable framework that enables the integration of advanced functionalities such as Firmware Over-the-Air (FOTA). Its standardized approach ensures compatibility, interoperability, and efficient management of resources across diverse Electronic Control Units (ECUs) \cite{AUTOSAR_Architecture}.

Within the AUTOSAR framework, the Memory Stack (MemStack) plays a critical role in managing memory resources, such as Flash and EEPROM, which are essential for supporting firmware updates. The MemStack offers the following key services:

\begin{enumerate} \item \textbf{Abstraction Layer:} This layer facilitates access to memory devices by abstracting hardware-specific details, enabling seamless communication between software components and memory hardware \cite{AUTOSAR_Specs}. \item \textbf{Non-Volatile Memory Services:} These services manage data storage and retrieval in non-volatile memory during firmware updates, ensuring the persistence of critical data even after system resets or power failures \cite{Memory_Management}. \item \textbf{Data Integrity Mechanisms:} The MemStack incorporates mechanisms to verify the accuracy and reliability of data stored in memory, which is particularly important during operations like firmware flashing to prevent corruption and ensure system stability \cite{AUTOSAR_MemStack}. \end{enumerate}

AUTOSAR's layered software architecture provides a structured framework that seamlessly integrates with FOTA systems. By delineating clear interfaces and defining services for memory management, vehicle network communication, and diagnostics, AUTOSAR ensures that firmware updates can be executed efficiently without disrupting normal vehicle operations \cite{AUTOSAR_Software_Layers}. This structured approach addresses the growing complexity of modern vehicles, where software updates are critical for maintaining optimal performance and ensuring security against emerging threats.
}

\subsection{Overview of Unified Diagnostic Services (UDS)}

{In the domain of automotive embedded systems, vehicles often consist of electronic control units (ECUs) designed by multiple manufacturers. This diversity necessitates a standardized diagnostic communication protocol to ensure interoperability across different brands and Original Equipment Manufacturers (OEMs). The Unified Diagnostic Services (UDS) protocol, defined by the ISO 14229 standard, addresses this challenge by providing a universal framework for diagnostic communication \cite{ISO14229_Standard}.

The term "Unified" in UDS signifies its global applicability, enabling diagnostic and programming services across various OEMs, while "Diagnostics" encompasses activities such as fault detection, sensor calibration, and ECU reprogramming. UDS is predominantly used for off-board diagnostics during vehicle maintenance and servicing. In these scenarios, a diagnostic tester (client) interfaces with the vehicle’s ECUs (servers) through the On-Board Diagnostics (OBD) interface. This interaction allows the tester to send requests and receive results, facilitating tasks like error code retrieval and software updates \cite{UDS_Automotive_Applications}.

Operating within the application and session layers of the Open Systems Interconnection (OSI) model, UDS is independent of the physical communication layer, making it adaptable to widely-used automotive communication protocols such as CAN, LIN, FlexRay, and Ethernet. This versatility ensures compatibility and standardization, allowing manufacturers to streamline diagnostic operations across diverse vehicle models \cite{UDS_OSI_Model}.

UDS’s significance lies in its ability to unify diagnostic practices across the automotive industry, reducing complexity and enhancing reliability in diagnostic and maintenance activities.
}

\subsection{Communication Protocols in FOTA}

{Effective communication between Electronic Control Units (ECUs) is a cornerstone of successful Firmware Over-the-Air (FOTA) implementations. Among the various communication protocols available, the Controller Area Network (CAN) protocol has become the industry standard in automotive systems due to its robustness, real-time performance, and fault tolerance \cite{CAN_Overview}. CAN facilitates reliable data exchange between the master and target ECUs during the update process, ensuring firmware updates are delivered accurately and promptly.

The CAN protocol operates on a priority-based arbitration mechanism, where messages with higher priority identifiers preempt lower-priority messages to gain bus access. This system ensures that critical update commands are transmitted without delay, even in congested networks \cite{CAN_Arbitration}. Furthermore, CAN incorporates robust error detection and correction features, such as cyclic redundancy checks (CRC), to maintain data integrity during transmission \cite{CAN_ErrorDetection}.

For high-speed data transfer, particularly between the ESP8266 module and the master ECU, the Serial Peripheral Interface (SPI) protocol is employed. SPI offers a full-duplex communication channel, allowing simultaneous bidirectional data transfer. This capability enables rapid transmission of firmware packages with minimal latency, making it ideal for handling large updates where efficiency is crucial to minimize overall update duration \cite{SPI_Benefits}.

By integrating CAN and SPI protocols, the proposed FOTA system achieves a balance of reliability and efficiency, meeting the stringent demands of modern automotive applications. These communication frameworks serve as the backbone of the system, ensuring seamless interaction between components and facilitating secure, efficient, and timely firmware updates \cite{FOTA_CommunicationProtocols}.}

\subsection{FOTA in Real-World Applications}

{The adoption of Firmware Over-the-Air (FOTA) systems has become a necessity in the modern automotive landscape due to the increasing complexity of software-driven functionalities and the need for frequent updates. Prominent automotive manufacturers, such as Tesla, have demonstrated the transformative impact of FOTA in maintaining vehicle performance and security. Tesla’s over-the-air update system has set a benchmark in the industry by enabling the deployment of new features, performance enhancements, and security patches without requiring service center visits \cite{Tesla_FOTA_Impact}.

FOTA systems do not merely enhance the user experience but also provide significant cost savings for manufacturers. By reducing the frequency of warranty claims and eliminating the logistical expenses associated with physical recalls, these systems offer a more streamlined approach to software maintenance. Additionally, FOTA enables manufacturers to collect and analyze diagnostic data, which facilitates predictive maintenance strategies. This capability enhances vehicle reliability and ensures that potential issues are identified and addressed proactively \cite{FOTA_Predictive_Maintenance}.

Moreover, FOTA contributes to the sustainability goals of the automotive industry by reducing the environmental impact of traditional recall processes. The reduced need for physical interventions translates to lower carbon emissions associated with transportation and servicing, aligning with global efforts to make the automotive sector more eco-friendly \cite{FOTA_Sustainability}.}

\subsection{Introducing Lane Keeping Assist (LKA)}

{Lane Keeping Assist (LKA) is a critical feature of Advanced Driver Assistance Systems (ADAS) aimed at enhancing road safety by maintaining vehicle alignment within its designated lane. This functionality is particularly valuable in mitigating unintentional lane departures, which are a leading cause of road accidents \cite{LaneKeepingSystems_Review}. In this project, LKA has been implemented as a test application for evaluating firmware updates on the STM32F401RE Nucleo board. The key aspects of the LKA system are outlined below:

\begin{itemize}
    \item \textbf{Functionality:}
    \begin{itemize}
        \item LKA systems actively intervene to prevent lane departure by applying steering corrections.
        \item Unlike Lane Departure Warning Systems (LDWS), which only alert the driver, LKA takes corrective action when the vehicle begins to drift unintentionally \cite{LKA_Vs_LDWS}.
    \end{itemize}
    
    \item \textbf{Operation:}
    \begin{itemize}
        \item A front-facing camera captures road markings in real-time, ensuring accurate lane detection.
        \item The video feed is processed using a Python-based artificial intelligence (AI) model hosted on an external laptop ECU.
        \item The model calculates lane deviations and transmits this data to the STM32F401RE Nucleo board through USART communication.
        \item Corrective steering actions are executed by a DC motor, which is connected to the vehicle's steering mechanism via a Cytron motor driver \cite{LKA_Implementation}.
    \end{itemize}
    
    \item \textbf{Levels of Assistance:}
    \begin{itemize}
        \item Basic LKA systems offer minimal steering corrections to prevent lane departure.
        \item Advanced systems, such as Lane Following Assist (LFA), actively center the vehicle within the lane and may integrate with Adaptive Cruise Control (ACC) for enhanced driver convenience and safety \cite{LFA_Integration}.
    \end{itemize}
\end{itemize}
}
The integration of LKA as a test application not only demonstrates the capabilities of the firmware update system but also highlights the potential for enhanced vehicle safety through real-time corrective interventions.

\subsubsection{Integration with Vehicle Dynamics Control}
{Lane Keeping Assist (LKA) systems are intricately integrated with vehicle dynamics control mechanisms, utilizing electric power steering systems to maintain precise lane positioning across diverse driving conditions. Research studies such as \cite{LKA_ISO11270} emphasize the capability of LKA systems to comply with ISO11270 standards, ensuring reliable performance even under varying lateral and longitudinal velocities. This integration with real-time processing and electric power steering systems significantly enhances the adaptability and reliability of LKA in modern vehicles.
  \begin{itemize}
    \item \textbf{Key Benefits:}
    \begin{itemize}
        \item Reduces unintentional lane departures, a leading cause of traffic accidents.
        \item Enhances road safety by reducing driver workload.
        \item Provides a framework for testing and validating Firmware Over-the-Air (FOTA) updates for automotive systems.
    \end{itemize}

    \item \textbf{Terminology and Scope:}
    \begin{itemize}
        \item LKA focuses on maintaining lane discipline by correcting steering deviations.
        \item Terminology overlaps with related technologies such as Lane Keeping Aid, Lane Following Assist, and Active Lane Keeping.
    \end{itemize}
\end{itemize}
}

\subsubsection{Importance in ADAS}
Lane Keeping Assist (LKA) stands as a pivotal feature within Advanced Driver Assistance Systems (ADAS), showcasing the practical implementation of advanced control algorithms alongside real-time data processing. By maintaining lane discipline and assisting drivers, LKA serves as an essential test case for assessing the performance of embedded systems in autonomous vehicle contexts. Furthermore, it highlights the seamless integration of AI-driven decision-making processes with real-time motor control systems, offering a robust framework for enhancing vehicle safety and autonomy \cite{ADAS_LKA}.

\section{Baseline of Our Study}
{This study builds upon established methodologies and frameworks for Firmware Over-the-Air (FOTA) systems, addressing critical limitations identified in prior research. Current FOTA implementations often assume predefined vehicle memory structures, fixed communication protocols, and static security policies. While such assumptions simplify initial system design, they constrain adaptability to the rapidly evolving demands of the automotive industry and fail to address user-centric concerns such as update time, bandwidth efficiency, and overall system security.

Previous studies, such as those conducted by \cite{Traditional_FOTA_Study}, have focused on traditional FOTA systems reliant on full firmware transfers, where the entire firmware image is updated regardless of the scope of changes. This approach, while straightforward, is bandwidth-intensive, time-consuming, and inefficient. Furthermore, existing solutions often utilize static security frameworks, primarily relying on fixed authentication protocols. While effective against known vulnerabilities, these frameworks lack adaptability to address emerging cyber threats effectively.

Our research addresses these limitations by adopting a novel approach that integrates delta updating, as highlighted in studies like \cite{Delta_Update_Efficiency}. Delta updating transmits only the differences between firmware versions, significantly reducing data transmission size and flashing time. This methodology improves the overall user experience and minimizes bandwidth usage, a key requirement in modern automotive software management.

Additionally, the use of the AUTOSAR architecture, as supported by \cite{AUTOSAR_Foundation}, provides a modular and scalable framework for memory management. AUTOSAR's layered structure ensures adaptability to various vehicle configurations without extensive modifications, making it an ideal solution for implementing advanced FOTA systems.

Security remains a critical focus. Unlike traditional FOTA systems employing static authentication protocols, our study emphasizes the use of UDS 0x27, which provides dynamic authentication mechanisms resistant to tampering and adaptable to emerging threats \cite{UDS_Protocol_Security}. This dynamic approach ensures that firmware updates remain secure, reliable, and robust in the face of evolving cyber challenges.

Efficient communication between system components is another cornerstone of our study. Leveraging the CAN protocol for communication between the master and target ECUs, as well as SPI for data transfer between the ESP8266 module and the master ECU, ensures real-time and reliable data transmission. These communication frameworks, as demonstrated in \cite{CAN_SPI_Integration}, address the increasing complexity of modern vehicles and enhance system performance.

Unlike prior studies that rely heavily on static configurations, our research adopts a dynamic and user-centric perspective. By integrating delta updating, modular architecture, and adaptive security measures, our proposed solution aligns with the evolving demands of both manufacturers and end-users. This adaptability ensures the relevance and robustness of our system in the face of rapid technological advancements.

Through this innovative approach, we aim to establish a robust baseline for future FOTA systems, combining efficiency, scalability, and security to meet the challenges of modern automotive software management.
}

\noindent  
\section{Background}
\label{sec:BG}
This section delves into the theoretical foundations, methodologies, and technologies that underpin the design and implementation of the Firmware Over-the-Air (FOTA) system presented in this project. By addressing secure communication, efficient memory management, and advanced automotive system architectures, this study lays the groundwork for a robust and scalable solution.

\subsection{AUTOSAR and Memory Management in FOTA Systems}

The AUTOSAR (Automotive Open System Architecture) standard has become indispensable for scalable and modular software development in the automotive sector. Its standardized platform facilitates seamless integration between software and hardware, promoting interoperability across various Electronic Control Units (ECUs). As emphasized by \cite{AUTOSAR_Advancements}, AUTOSAR decouples application-level functions from low-level hardware operations, enabling developers to reuse software components efficiently.

Memory management within AUTOSAR adheres to stringent specifications designed to ensure secure and efficient firmware handling. The AUTOSAR Memory Access Specification outlines protocols for reading, writing, and erasing memory regions. These mechanisms are particularly critical for FOTA systems, as they underpin the implementation of delta updates—an approach that transmits only the differences between firmware versions, significantly reducing data size and flashing time \cite{Delta_Updates_AUTOSAR}.

\subsection{Secure Communication Protocols in FOTA}

Security is a foundational aspect of FOTA systems, given the risks associated with wireless updates. The UDS (Unified Diagnostic Services) protocol, particularly the 0x27 service (Seed and Key), is widely adopted for authenticating update requests and mitigating unauthorized access. This challenge-response mechanism ensures that only authenticated entities can execute firmware updates, safeguarding the system against tampering \cite{UDS_Authentication}.

Emerging advancements, such as UDS 0x29, have introduced enhanced cryptographic measures to counteract evolving cybersecurity threats. Studies on automotive cybersecurity suggest that dynamic key generation and advanced cryptographic techniques can significantly improve resilience against replay and man-in-the-middle attacks, making them vital for modern FOTA implementations \cite{Automotive_Security_Protocol}.

\subsection{The Role of CAN in Automotive Communication}

The Controller Area Network (CAN) protocol is a cornerstone of automotive communication systems, renowned for its robustness, fault tolerance, and real-time capabilities. CAN's priority-based arbitration and error detection mechanisms make it ideal for safety-critical applications like FOTA, ensuring timely and reliable communication \cite{CAN_RealTime}.

In this project, CAN serves as the primary communication medium between the master and target ECUs. Its deterministic nature facilitates the efficient transmission of delta updates, while its error-checking features enhance data integrity. Research highlights that CAN maintains performance even in noisy environments, validating its suitability for demanding automotive applications \cite{CAN_Applications}.

\subsection{Delta Updating: Efficiency and Scalability}

Delta updating has revolutionized firmware management by transmitting only the differences between successive firmware versions. This approach minimizes data transfer requirements, reduces update time, and conserves bandwidth—a critical factor in resource-constrained environments \cite{Delta_Update_Benefits}.

Studies on patch-based updating techniques demonstrate significant efficiency gains in distributed systems. By integrating delta updating within the AUTOSAR framework, this project capitalizes on these benefits while ensuring compliance with industry standards. This integration not only optimizes data usage but also accelerates the deployment of updates across diverse vehicle configurations \cite{Patch_Update_Study}.

\subsection{FOTA and Advanced Automotive Applications}

The adoption of FOTA systems in modern vehicles has revolutionized automotive maintenance, enabling rapid software deployment to ensure security and functionality. Applications like lane-keeping assist systems and autonomous driving heavily depend on timely updates for optimal performance \cite{FOTA_Applications}.

Recent advancements in FOTA technology have showcased its potential to enhance user experience and operational efficiency. By integrating delta updating, secure communication protocols, and robust memory management, this project addresses the limitations of traditional update methods, paving the way for future innovations in automotive software management \cite{Future_FOTA}.
\clearemptydoublepage

\chapter{Proposed Solution}
\label{chap:PS}

\section{Memory Architecture}
\label{subsec:MS}
{In this project, we propose a memory sectoring strategy for single-bank Flash memory in STM32F4xx microcontrollers. The proposed layout divides the memory into three key regions, each serving a distinct purpose:
\begin{enumerate}
\item \textbf{Boot Manager:} the Boot Manager is responsible for handling the initial system setup. It determines the appropriate actions during startup, such as launching the bootloader or the main application.
\item \textbf{Bootloader:} the Bootloader manages firmware updates. It ensures secure and reliable programming of the application during updates, acting as a crucial intermediary for FOTA processes.
\item \textbf{Application and Bootloader Updater:} both share the same memory region. The Bootloader Updater is responsible for updating the bootloader itself. 
\end{enumerate}
By leveraging sectoring, this memory architecture balances flexibility, efficiency, and safety, making it highly suitable for automotive systems with stringent reliability and performance requirements.}

\section{CAN Protocol for ECU Communication}
\label{subsec:CP}
{We propose using the CAN protocol as the communication interface between the target and master ECUs. This choice aligns with the ISO 11898 standard, which specifies the CAN protocol as a reliable and robust solution for real-time communication in automotive systems. By leveraging CAN, we ensure compatibility with industry standards and exploit its inherent strengths. 

CAN offers several benefits, including high reliability due to its built-in error detection and handling mechanisms, which ensure data integrity during transmission. Its fault confinement capabilities prevent network disruptions caused by malfunctioning nodes, while prioritized message arbitration supports real-time communication. Additionally, CAN provides scalability, allowing multiple ECUs to coexist seamlessly on the same network and robustness through differential signaling that minimizes electromagnetic interference (EMI). The protocol’s simplicity and widespread adoption contribute to its cost-effectiveness and its standardization under ISO 11898 guarantees compatibility across a wide range of devices and manufacturers.
}

\section{Security}
\subsubsection{UDS Protocol: 0x27 Security Access Service}
\label{subsec:uds_0x27_security_access}
As a part of the Firmware Over-the-Air (FOTA) update system, the 0x27 Security Access service from the Unified Diagnostic Services (UDS) protocol has been chosen to secure critical communication between the tester and the Electronic Control Unit (ECU). This service is essential for ensuring that firmware updates and parameter modifications are performed securely and exclusively by authorized entities.

The 0x27 Security Access service operates by establishing an authentication mechanism using a challenge-response model. This approach protects the system from unauthorized access and ensures the integrity of sensitive ECU operations. The ECU generates a unique seed and expects a corresponding key, calculated using a predefined algorithm, to grant access to secure functions. This authentication prevents malicious actors from tampering with the system, thereby safeguarding the vehicle’s safety and reliability.

Integrating this service into our solution aligns with ISO 14229 standards, ensuring compatibility with existing automotive diagnostic protocols and enhancing the overall security of the FOTA system.

\section{Memory Stack}
This section presents our proposed solution for robust and efficient management of persistent data within our embedded system. Recognizing the critical need for reliable data storage in resource-constrained automotive applications, we designed a memory management strategy leveraging the AUTOSAR Memory Stack and tailored it for the STM32F401RE Nucleo platform. This approach aims to address key challenges such as limited memory, data integrity risks from power loss or interference, and the requirement for deterministic behavior, while also facilitating efficient storage and retrieval of configuration parameters and enabling delta updates.

Our proposed solution centers on the integration of the AUTOSAR Memory Stack. This choice provides a modular architecture and standardized interfaces, which are expected to simplify development, enhance portability, and ensure compliance with automotive standards. We \textit{propose} to adapt this framework specifically for the STM32F401RE Nucleo board, implementing the necessary adaptations to enable reliable management of persistent data. This implementation will ensure efficient data storage and retrieval, while also supporting the delta update mechanism crucial for our project's requirements.
\label{subsec:Adaptation of the AUTOSAR Memory Stack for STM32F401RE }
To achieve seamless integration with the STM32F401RE's Flash memory controller, we propose to customize the Flash driver (Fls). This customization will enable reliable read, write, and erase operations essential for persistent storage. The proposed key adaptations include:

\begin{itemize}
    \item \textbf{Read Operation:} We plan to implement functions that will allow access to specific Flash memory addresses.
    \item \textbf{Write Operation:} Our proposed write operation will incorporate proper handling of Flash programming sequences, including sector erasure and data writing procedures. This will be designed to minimize write times and maximize Flash endurance.
    \item \textbf{Erase Operation:} We propose an efficient sector erase process that will support future updates and enable the reclamation of memory for new data.
    \item \textbf{Block Configuration:} To support efficient delta updates, we propose dividing the program section of the Flash memory into 128 blocks, each with a 1 KB capacity. This configuration was chosen to accommodate the expected size of our configuration data and to provide sufficient granularity for delta updates. This allows us to update only the changed blocks, minimizing write cycles and extending Flash memory lifespan. 
\end{itemize}

The customized Flash driver will be integrated into the AUTOSAR Basic Software (BSW) through the Memory Abstraction Interface (MemIf). This abstraction is intended to provide the Non-Volatile Memory Manager (NvM) with transparent access to Flash memory. The NvM module will then manage read/write operations and communicate with application software via the AUTOSAR Runtime Environment (RTE), promoting modularity and scalability within the system.

 \section{Delta Update Mechanism using CRC }
 To optimize NVM write endurance and minimize update times, we propose a delta update mechanism employing Cyclic Redundancy Checks (CRCs). This method addresses the inherent limitations of Flash memory regarding write cycles and update overhead, crucial in resource-constrained embedded systems.

The proposed approach leverages CRC checksums calculated for each 1KB NVM block. These CRCs are stored alongside the corresponding data, enabling integrity verification. During an update, a differential analysis (performed off-chip) identifies changes between the current and new data versions, generating a delta represented as (offset, length, data) tuples.

The on-chip update procedure comprises the following steps:(1) Receiving the delta information; (2) Reading the target NVM block; (3) Applying the delta tuples to modify the block data; (4) Calculating a new CRC for the updated block; and (5) Writing the modified block, including the updated CRC, to NVM. Post-update, the calculated CRC is compared with the stored CRC to validate data integrity. A mismatch triggers an error and potentially a rollback procedure.

This delta update strategy offers significant advantages: (1) Reduced write cycles, extending Flash lifespan; (2) Faster update times due to minimized data transfer; (3) Enhanced data integrity through CRC verification; and (4) Efficient storage by transmitting only differential data. This approach, integrated with the AUTOSAR Memory Stack, provides a robust and efficient solution for persistent data management, crucial for reliable automotive embedded systems.

 \section{FreeRTOS}
To manage the concurrent operations related to NVM access, data communication, and other system functionalities, we propose the integration of FreeRTOS, a real-time operating system (RTOS), into our embedded system. This choice addresses the need for deterministic task scheduling, efficient resource management, and streamlined inter-task communication, crucial for achieving reliable and predictable system behavior.

FreeRTOS provides a preemptive, priority-based scheduler that allows us to assign different priorities to various tasks, ensuring that time-critical operations, such as NVM read/write requests and communication handling, are executed promptly. This deterministic scheduling is essential for meeting real-time constraints within the automotive environment. 

Specifically, we propose to organize the following functionalities as separate FreeRTOS tasks:

\begin{itemize}
    \item \textbf{NVM Management Tasks:} Responsible for handling all NVM read/write requests, coordinating access to the NvM module, and ensuring data consistency.
    \item \textbf{Communication Tasks:} Manages the communication interface, receiving and transmitting data to/from external devices or other ECUs.
    \item \textbf{Application Tasks:} Implement the core application logic of the system.
\end{itemize}

The integration of FreeRTOS provides several advantages: (1) Deterministic task scheduling, ensuring timely execution of critical operations; (2) Efficient resource management, optimizing system performance; (3) Simplified inter-task communication, streamlining software development; and (4) Improved system modularity, enhancing maintainability and scalability. By leveraging FreeRTOS, we aim to achieve a robust, efficient, and well-structured embedded system capable of handling concurrent operations related to NVM management and communication.

\section{Lane Keeping Assist (LKA)}
\label{subsec:LKA}

The Lane Keeping Assist (LKA) application in this project focuses on evaluating and enhancing the performance of the PID controller in a lane-centering system. Instead of utilizing a real-world prototype vehicle, the testing is conducted using pre-recorded video streams that simulate real-world driving scenarios. These videos are processed using a Python script to calculate the deviation of the car from the lane center, which is then transmitted to the STM32 Nucleo board to adjust the steering wheel accordingly.
 The process can be summarized as follows:

\begin{itemize}
    \item \textbf{System Architecture:}
    \begin{itemize}
        \item The system comprises a DC motor connected to a steering wheel, controlled by a Cytron motor driver, which interfaces with an STM32F401RE Nucleo board.
        \item An incremental encoder provides feedback on the steering wheel's position.
        \item Pre-recorded videos of street scenarios, captured using a front-facing camera, are fed into a laptop that simulates the electronic control unit (ECU).
    \end{itemize}

    \item \textbf{Data Processing:}
    \begin{itemize}
        \item A Python script running on the laptop processes the video streams using an AI model to determine the vehicle's deviation from the lane center.
        \item The deviation values are transmitted to the Nucleo board via USART communication.
    \end{itemize}

    \item \textbf{Control Mechanism:}
    \begin{itemize}
        \item The Nucleo board calculates the appropriate steering adjustments using a PID controller based on the received deviation values.
        \item The PID parameters (Kp, Ki, Kd) can be updated through FOTA updates to optimize performance or test alternative control mechanisms.
    \end{itemize}

    \item \textbf{FOTA Integration:}
    \begin{itemize}
        \item The system allows for seamless updates to the firmware, enabling the deployment of new PID parameters or entirely new control algorithms without physical intervention.
        \item This ensures that the system remains adaptable and capable of continuous improvement.
    \end{itemize}

    \item \textbf{Testing Framework:}
    \begin{itemize}
        \item The solution is tested in a controlled environment by feeding the pre-recorded videos to the Python script and monitoring the system's response.
        \item The performance is evaluated based on the ability of the steering system to maintain lane centering.
    \end{itemize}
\end{itemize}

This approach minimizes the complexities and risks associated with real-world testing while providing a robust framework for validating the proposed enhancements to the LKA system.

\clearemptydoublepage

\chapter{Implementation}
\label{chap: IM}
\section{Memory Architecture}
\subsection{Overview of Memory Architecture}
Memory is the backbone of modern automotive systems, serving as a critical enabler for advancements like Firmware Over-the-Air (FOTA) updates. In FOTA systems, memory’s role extends far beyond simple data storage—it is the foundation for ensuring the reliability, efficiency, and security of the entire update process. As vehicles become increasingly connected and reliant on software-driven features, the importance of memory in handling these demands cannot be overstated.

Memory is responsible for storing firmware images, managing delta updates, and enabling rollback mechanisms in case of update failures. Its architecture must be designed to meet the stringent requirements of automotive standards such as AUTOSAR, which prioritize safety, modularity, and fault tolerance. Additionally, memory must support high read/write speeds, wear-leveling techniques, and secure storage to protect against data corruption and unauthorized access.

By ensuring seamless operation during updates, memory not only minimizes vehicle downtime but also safeguards critical systems from potential risks. Its strategic role in maintaining the integrity and functionality of software updates highlights why memory is a pivotal element in the success of FOTA technology and the future of the automotive industry.

\subsubsection{Embedded Flash Memory Interface }
\label{subsec:flash_memory_interface}
The Flash memory interface manages CPU AHB I-Code and D-Code accesses to the Flash memory. It implements the erase and program Flash memory operations and the read and write protection mechanisms. The Flash memory interface accelerates code execution with a system of instruction prefetch and cache lines \cite{STM32F4_Reference_Manual}.

\subsubsection{Main Features}

\begin{itemize}
    \item \textbf{Flash memory read operations:} Enables efficient reading from the Flash memory.
    \item \textbf{Flash memory program/erase operations:} Handles programming and erasing of Flash memory sectors.
    \item \textbf{Read/write protections:} Ensures data integrity and security by providing mechanisms to protect against unauthorized access.
    \item \textbf{Prefetch on I-Code:} Speeds up instruction fetch by preloading instructions into the cache.
    \item \textbf{64 cache lines of 128 bits on I-Code:} Optimizes instruction execution by caching frequently accessed instructions.
    \item \textbf{8 cache lines of 128 bits on D-Code:} Enhances data access efficiency by caching frequently used data.
\end{itemize}
\begin{figure}[h]
    \centering
    \includegraphics[width=1.0\linewidth]{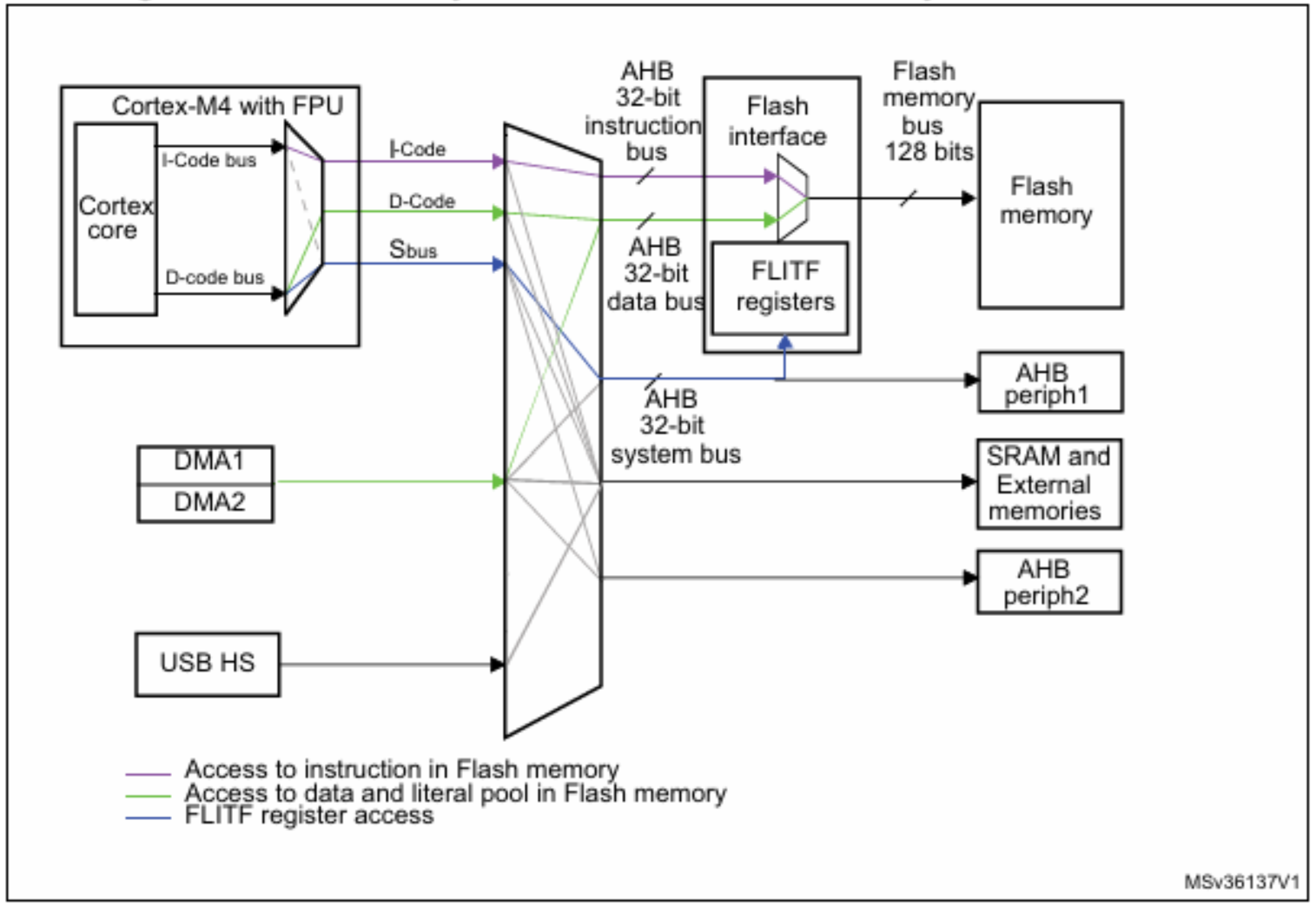}
    \caption{Flash Memory Interface Connection Inside System Architecture, \cite{STM32F4_Reference_Manual}}
    \label{fig:enter-label}
\end{figure}

\newpage
\subsubsection{Flash Memory Features}
\label{subsubsec:flash_memory_features}

The Flash memory in STM32F4xx microcontrollers offers several advanced features and capabilities as listed in \cite{STM32F4_Reference_Manual}:

\begin{itemize}
    \item \textbf{Capacity:} Up to 512 Kbytes.
    \item \textbf{Data Operations:} 128 bits wide data read, with byte, half-word, word, and double word write operations.
    \item \textbf{Erase Operations:} Supports sector and mass erase functionalities.
    \item \textbf{Memory Organization:} The Flash memory is organized as follows:
    \begin{itemize}
        \item A main memory block divided into 4 sectors of 16 Kbytes, 1 sector of 64 Kbytes, and 3 sectors of 128 Kbytes.
        \item System memory from which the device boots in System memory boot mode.
        \item 512 OTP (one-time programmable) bytes for user data. The OTP area contains 16 additional bytes used to lock the corresponding OTP data block.
        \item Option bytes to configure read and write protection, BOR level, watchdog software/hardware, and reset when the device is in Standby or Stop mode.
    \end{itemize}
    \item \textbf{Low-Power Modes:} Flash memory supports operation in low-power modes.
\end{itemize}
\begin{figure}[h]
    \centering
    \includegraphics[width=1.0\linewidth]{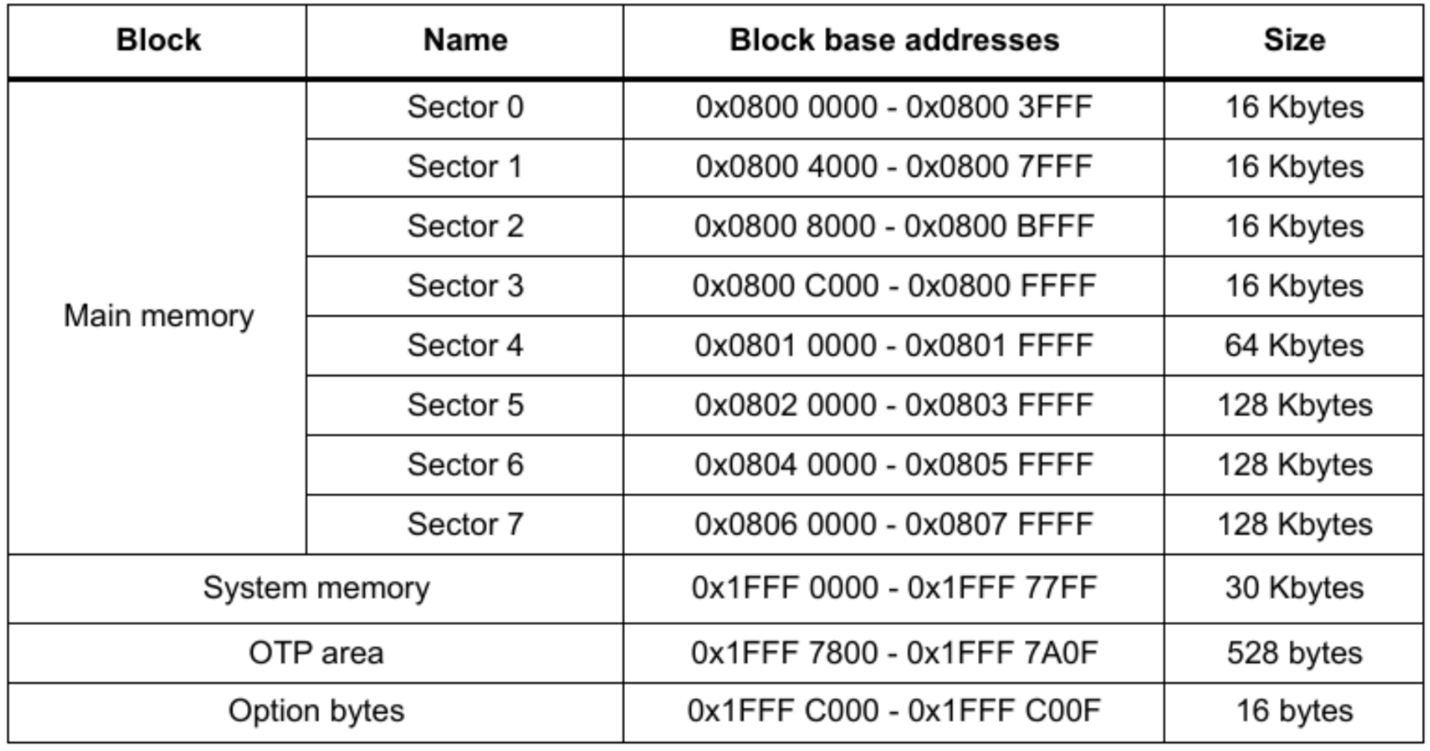}
    \caption{Flash Module Organization, \cite{STM32F4_Reference_Manual}}
    \label{fig:enter-label}
\end{figure}
For any Flash memory program operation (erase or program), the CPU clock frequency (HCLK) must be at least 1 MHz. The contents of the Flash memory are not guaranteed if a device reset occurs during a Flash memory operation.

Any attempt to read the Flash memory on STM32F4xx while it is being written or erased, causes the bus to stall. Read operations are processed correctly once the program operation has been completed. This means that code or data fetches cannot be performed while a write/erase operation is ongoing.

\subsection{Memory Sectoring in Single-Bank Flash}
Memory Sectoring in Single-Bank Flash
To implement memory sectoring, a linker script is used to define memory regions and allocate specific sectors for designated purposes. The linker script provides precise control over the memory layout by mapping specific addresses to functional sections of the firmware. 
Using a linker script ensures that each memory sector is utilized efficiently and that critical sections are isolated for security and reliability. The script also facilitates flexibility, allowing developers to modify memory allocations as system requirements evolve. Sectoring enables streamlined update mechanisms, ensuring that unused or reserved sections of memory are leveraged optimally without affecting operational firmware.

\subsection{Memory Layout}
\label{subsec:memory_layout}

The memory layout in STM32F4xx boards is carefully organized to accommodate various firmware components and functionalities. In the proposed configuration:

\begin{itemize}
    \item \textbf{Boot Manager:} 
    The first sector, \textbf{sector 0}, comprising 64 KB, is allocated for the Boot Manager. This component handles the initial system setup and determines the appropriate actions during startup, such as launching the bootloader or application.

    \item \textbf{Bootloader:} 
    The fifth sector, \textbf{sector 4}, comprising 64 KB, is designated for the Bootloader. This component manages firmware updates and ensures secure and reliable programming of the application.

    \item \textbf{Application and Bootloader Updater:} 
    The sixth sector, \textbf{sector 5}, comprising 384 KB, is shared between the Application and the Bootloader Updater. The Bootloader Updater is responsible for updating the Bootloader itself.
\end{itemize}
\begin{figure}[h]
    \centering
    \includegraphics[width=1.0\linewidth]{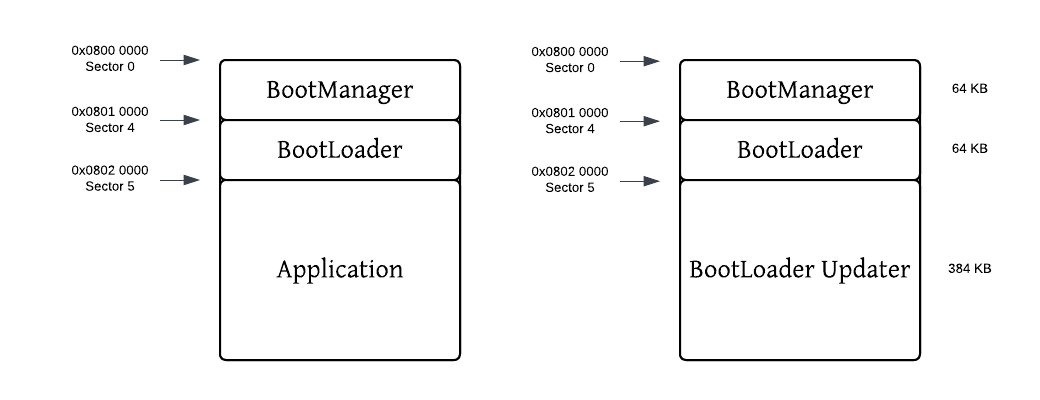}
    \caption{Memory Layout}
    \label{fig:enter-label}
\end{figure}
\newpage
This structured memory layout ensures that critical system components are isolated while maximizing memory utilization. The shared memory region between the Application and the Bootloader Updater highlights the flexibility of sectoring and linker scripts in optimizing space without compromising functionality.

\subsection{Memory Module Implementation}
\label{subsec:memory_module}

The memory module in the project is designed to leverage the STM32F4xx microcontroller’s Flash memory, offering essential functions for writing, reading, and erasing operations. These functionalities are implemented with a focus on efficiency and compatibility with the microcontroller's architecture.

\subsubsection{Memory Module Main Functions}

\begin{lstlisting}[language=C]
    uint8_t Perform_Flash_Erase(uint8_t Sector_Number, uint8_t Number_Of_Sectors);
\end{lstlisting}
    \begin{itemize}
        \item \textbf{Sector\_Number:} The starting sector number to erase.
        \item \textbf{Number\_Of\_Sectors:} The number of sectors to erase starting from Sector\_Number.
    \end{itemize}
This function erases one or more sectors in Flash memory starting from the specified sector number. The function unlocks the Flash memory, performs the erase operation, and locks it again to ensure data integrity.

\begin{lstlisting}[language=C]
    uint8_t Flash_Memory_Write_Payload(uint8_t *Host_Payload, uint32_t Payload_Start_Address, uint16_t Payload_Len);
\end{lstlisting}
    \begin{itemize}
        \item \textbf{Host\_Payload:} Pointer to the data payload to be written.
        \item \textbf{Payload\_Start\_Address:} The Flash memory address where the payload writing begins.
        \item \textbf{Payload\_Len:} Length of the payload in bytes.
    \end{itemize}
This function writes a payload of data into the Flash memory at the specified starting address. The function writes each byte sequentially and handles errors to ensure successful completion. It unlocks the Flash memory, programs the payload, and locks the memory afterward to maintain security and stability.

\subsection{Error Handling}

The memory module incorporates robust error-handling mechanisms. If an operation fails, an error code is returned to the application. The module utilizes the STM32F4’s built-in error-handling features, providing detailed diagnostics to identify and resolve issues effectively. For more information, refer to the STM32F4 reference manual.

\subsection{Boot Manager}
A Boot Manager is a critical component of embedded systems that operates as the first decision-making entity during system startup or reset. Unlike a bootloader, which focuses on initializing the system and handling firmware updates, the Boot Manager’s primary role is to manage the overall control flow and determine the appropriate operational mode for the system. It serves as a high-level coordinator, ensuring that the system transitions smoothly between different states, such as launching the bootloader, executing the main application, or initiating recovery procedures.

In systems with complex requirements, such as dual-boot configurations or systems with both a bootloader and application, the Boot Manager’s role becomes indispensable. It ensures that the system selects the correct path based on predefined conditions, flags, or user inputs.

\subsubsection{Boot Manager Behavior}
\label{subsec:boot_manager_behavior}

The Boot Manager operates within a dedicated memory segment and performs several key tasks to ensure reliable system operation:

\begin{enumerate}
    \item \textbf{System Initialization:} 
    The Boot Manager initializes basic system peripherals and resources required for its operation. This includes setting up communication interfaces and reading any critical system flags or registers.
    
    \item \textbf{Decision-Making Logic:} 
    Based on system conditions and inputs, the Boot Manager determines the next operational state:
    \begin{itemize}
        \item Launching the Bootloader for firmware updates.
        \item Jumping to the main application for normal operation.
        \item Launching the Bootloader Updater for Bootloader updates.
    \end{itemize}
    
    \item \textbf{Flag Management:} 
    The Boot Manager reads and updates system flags stored in backup registers or non-volatile memory. These flags indicate the desired system state, such as entering the Bootloader for updates or jumping directly to the application.
    
    \item \textbf{Error Handling and Recovery:} 
    If the system encounters a critical error or an invalid state, the Boot Manager initiates recovery actions. This may include logging errors, resetting the system, or triggering fallback mechanisms.
    
    \item \textbf{Control Handoff:} 
    Once the appropriate decision is made, the Boot Manager transfers control to the selected component – Bootloader, Bootloader Updater, or Application – to ensure seamless system operation.
\end{enumerate}

\subsubsection{Boot Manager Memory Integration}
\label{subsec:boot_manager_memory_integration}

In the system’s memory architecture:

\begin{itemize}
    \item \textbf{Execution Priority:} 
    The Boot Manager resides in the first memory segment; specifically, at sector 0,  ensuring it is the first entity executed after startup or reset.
    
    \item \textbf{Component Interaction:} 
    It interacts with other components – the Bootloader, the Bootloader Updater, and the Application – by reading and updating system flags and control variables.
    
    \item \textbf{Memory Isolation:} 
    The Boot Manager’s isolated memory space ensures its operation remains independent of other system components, enhancing reliability.
\end{itemize}

\subsubsection{Boot Manager Main Function}
Below is an overview of the boot manager main function, App\_Logic():
\begin{lstlisting}[language=C, caption={App\_Logic() Function}, label={lst:app_logic}]
void App_Logic(){

    // Toggling LED
    for (uint8_t var = 0; var < 3; ++var) {
        HAL_GPIO_WritePin(LD2_GPIO_Port, LD2_Pin, GPIO_PIN_SET);
        HAL_Delay(200);
        HAL_GPIO_WritePin(LD2_GPIO_Port, LD2_Pin, GPIO_PIN_RESET);
        HAL_Delay(200);
    }

    /*
     * Calculate Application Integrity
     */
    uint32_t app_no_bytes = atoi((const char *)(APP_NO_OF_BYTES_START_ADDRESS));
    uint32_t stored_crc = *((uint32_t *)(APP_CRC_START_ADDRESS));
    uint32_t calculated_crc = CalculateCRC((const uint8_t *)APP_BINARY_START_ADDRESS, app_no_bytes);
    uint8_t Application_Integrity_result = CRCCompare(calculated_crc, stored_crc);

    /*
     * Reading Control Flags
     */
    uint8_t Application_Enter_flag = Read_RTC_backup_reg(APPLICATION_ENTER_FLAG_ADDRESS);
    uint8_t BL_Updater_Enter_flag = Read_RTC_backup_reg(BOOTLOADER_UPDATER_ENTER_FLAG_ADDRESS);

    /*
     * Branching Conditions
     */
    Application_Enter_flag = ENTER;
    if((Application_Integrity_result == SUCCEEDED) && (Application_Enter_flag == ENTER)) {
        // Jump to Application
        jump_to_Image_Address(APP_BINARY_START_ADDRESS);
    } else if (BL_Updater_Enter_flag == ENTER) {
        // Jump to Bootloader Updater
        jump_to_Image_Address(BOOTLOADER_UPDATER_BINARY_START_ADDRESS);
    } else {
        // Resetting flags (e.g., application flag = ENTER but its integrity is NOK)
        Write_RTC_backup_reg(APPLICATION_ENTER_FLAG_ADDRESS, N_ENTER);
        Write_RTC_backup_reg(BOOTLOADER_UPDATER_ENTER_FLAG_ADDRESS, N_ENTER);
        // Jump to Bootloader
        jump_to_Image_Address(BOOTLOADER_BINARY_START_ADDRESS);
    }
}
\end{lstlisting}

The \texttt{App\_Logic()} function operates as follows:

\begin{enumerate}
    \item \textbf{Toggle LED (Optional Debugging):} 
    Temporarily commented out, the LED toggling serves as a debug indicator during initialization.

    \item \textbf{Integrity Check:}
    \begin{itemize}
        \item Retrieve the stored CRC and number of bytes for the application.
        \item Calculate the CRC of the application binary and compare it with the stored CRC.
    \end{itemize}

    \item \textbf{Flag Reading:}
    \begin{itemize}
        \item Read the \texttt{Application\_Enter\_flag} and \texttt{BL\_Updater\_Enter\_flag} from RTC backup registers.
    \end{itemize}

    \item \textbf{Branching Conditions:}
    \begin{itemize}
        \item Execute the application if the integrity check succeeds and the application flag is set.
        \item Jump to the Bootloader Updater if the updater flag is set.
        \item Otherwise, reset the flags and jump to the Bootloader.
    \end{itemize}
\end{enumerate}

\subsection{Bootloader}
A bootloader is a fundamental piece of firmware that runs immediately after an embedded system is powered on or reset. Its primary role is to initialize the system and transfer control to the main application. In some systems, the bootloader also facilitates firmware updates, making it an essential component for maintaining and upgrading embedded devices.

For systems with multiple applications, such as dual-boot Electronic Control Units (ECUs), the bootloader decides which application to execute during each reset or startup. It may also allow downloading and verifying firmware updates, ensuring that the system is always running the latest and most secure software. The specific operations of a bootloader depend on the system’s architecture and its requirements.

In traditional setups, the bootloader operates during off-time and updates firmware via a diagnostic communication link, such as CAN or UART, providing a reliable mechanism for remote firmware management.

\subsubsection{Bootloader Behavior}
\label{subsec:bootloader_behavior}

Bootloaders, especially for non-automotive ECUs, consist of three main components:

\begin{enumerate}
    \item \textbf{Branching Code (Boot Manager):} 
    This component determines whether the bootloader or the main application should be executed. 
    \begin{itemize}
        \item In simple systems, this decision could rely on checking a GPIO pin or a specific flag.
        \item In more complex systems, the bootloader may load itself into memory, perform basic system checks, and validate system integrity before deciding the next step.
    \end{itemize}
    The branching code ensures that the system remains secure and operational under various conditions.

    \item \textbf{Application Code:} 
    The application code runs only after the branching code confirms that it is safe and appropriate to do so. 
    \begin{itemize}
        \item The application is designed to accept a command to re-enter the bootloader when necessary.
        \item Upon receiving such a command, it performs necessary cleanup operations and executes a soft reset, handing control back to the bootloader.
    \end{itemize}

    \item \textbf{Bootloader Code:} 
    The main responsibility of the bootloader code is to handle firmware updates.
    \begin{itemize}
        \item It initializes essential peripherals, such as the system clock and communication interfaces.
        \item Receives new firmware, verifies its integrity, and programs it into the Flash memory.
    \end{itemize}
    This ensures the system remains up-to-date and secure.
\end{enumerate}

\subsubsection{Traditional Bootloader Flow}
\label{subsec:traditional_bootloader_flow}

The typical bootloader flow involves:

\begin{itemize}
    \item \textbf{System Initialization:} 
    Initializing the system upon startup.

    \item \textbf{Decision-Making:} 
    Deciding whether to execute the bootloader or the main application.

    \item \textbf{Bootloader Mode:} 
    If in bootloader mode, managing the firmware update process by:
    \begin{itemize}
        \item Receiving the firmware via a communication protocol.
        \item Verifying the integrity of the firmware.
        \item Programming the firmware into Flash memory.
    \end{itemize}

    \item \textbf{Control Handoff:} 
    Transferring control to the main application after the update process or when deemed safe.
\end{itemize}

By structuring these processes, bootloaders ensure reliability, security, and flexibility in embedded systems. They are indispensable for systems requiring regular updates or maintaining operational integrity in complex, multi-application environments.

\begin{figure}[h]
    \centering
    \includegraphics[width=1.0\linewidth]{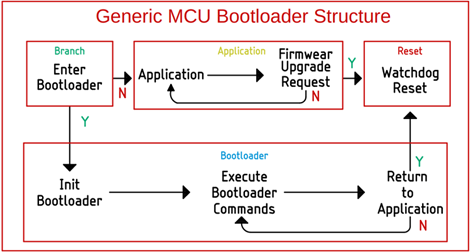}
    \caption{Generic MCU Bootloader Structure, \cite{Bootloader_Microcontroller}}
    \label{fig:enter-label}
\end{figure}

\subsubsection{Bootloader Memory Integration}
\label{subsec:bootloader_memory_integration}

In the system’s memory architecture:

\begin{itemize}
    \item \textbf{Dedicated Memory Segment:} 
    The Bootloader resides in a dedicated memory segment, isolating it from the Boot Manager and Application/Bootloader Updater; specifically, at sector 4. 

    \item \textbf{Integrity and Functionality:} 
    The Bootloader ensures the integrity and functionality of the firmware stored in the application region while leveraging the Boot Manager for high-level decision-making. It uses RTC backup registers to store update-related flags.

\end{itemize}

\subsubsection{Bootloader Main Function}
Below is an overview of the bootloader main function, BL\_CAN\_Fetch\_Host\_Command():
\begin{lstlisting}[language=C, caption={BL\_CAN\_Fetch\_Host\_Command() Function}, label={lst:bl_can_fetch_host_command}]
BL_Status BL_CAN_Fetch_Host_Command(void) {

    BL_Status Status = BL_NACK;
    HAL_StatusTypeDef HAL_Status = HAL_OK;
    uint8_t Data_Length = 0;

    memset(BL_Host_Buffer, 0, BL_HOST_BUFFER_RX_LENGTH);

    CAN_RxHeaderTypeDef RxHeader;
    uint8_t RxData[8];

    HAL_Status = HAL_CAN_GetRxMessage(&hcan1, CAN_RX_FIFO0, &RxHeader, RxData);

    if (HAL_Status != HAL_OK) {
        Status = BL_NACK;
    } else {
        // Process received CAN message
        memcpy(BL_Host_Buffer, RxData, RxHeader.DLC);
        Data_Length = RxHeader.DLC;

        switch (commands) {
            case CBL_GO_TO_ADDR_CMD:
                Leaving_App_Handler();
                Status = BL_OK;
                break;
            case CBL_FLASH_ERASE_CMD:
                Status = BL_OK;
                commands = 0x16;
                Bootloader_Erase_Flash(erasedata);
                break;
            case CBL_MEM_WRITE_CMD:
                Bootloader_Memory_Write(BL_Host_Buffer);
                Status = BL_OK;
                commands = 0x14;
                break;
            default:
                BL_Print_Message("Invalid command code received from host !! \r\n");
                break;
        }
    }
    return Status;
}
\end{lstlisting}

The \texttt{BL\_CAN\_Fetch\_Host\_Command()} function operates as follows:

\begin{enumerate}
    \item \textbf{Initialization:}
    \begin{itemize}
        \item A default \texttt{Status} is set to \texttt{BL\_NACK} (negative acknowledgment).
        \item Buffers are cleared to ensure no residual data interferes with command processing.
    \end{itemize}

    \item \textbf{Command Reception:}
    \begin{itemize}
        \item The function uses \texttt{HAL\_CAN\_GetRxMessage()} to fetch a message from the CAN receive FIFO.
        \item If the message is successfully retrieved, the payload (\texttt{RxData}) and its length (\texttt{RxHeader.DLC}) are copied into the \texttt{BL\_Host\_Buffer}.
    \end{itemize}

    \item \textbf{Command Parsing:}
    \begin{itemize}
        \item The received command is identified using a \texttt{switch-case} structure.
        \item Each case corresponds to a specific Bootloader operation, such as jumping to an application address, erasing memory, or writing to Flash memory.
    \end{itemize}

    \item \textbf{Command Execution:}
    \begin{itemize}
        \item \texttt{CBL\_FLASH\_ERASE\_CMD:} 
        Calls \texttt{Bootloader\_Erase\_Flash(erasedata)} to erase memory sectors.
        \item \texttt{CBL\_MEM\_WRITE\_CMD:} 
        Calls \texttt{Bootloader\_Memory\_Write(BL\_Host\_Buffer)} to write data to Flash memory.
         \item \texttt{CBL\_GO\_TO\_ADDR\_CMD:} 
        Calls \texttt{Leaving\_App\_Handler()} to transition to the Boot Manager.
    \end{itemize}

    \item \textbf{Error Handling:}
    \begin{itemize}
        \item If an invalid command is received, an error message is logged, and the function retains its \texttt{BL\_NACK} status.
    \end{itemize}

    \item \textbf{Status Return:}
    \begin{itemize}
        \item The function returns \texttt{BL\_OK} if the command was successfully processed or \texttt{BL\_NACK} in case of errors.
    \end{itemize}
\end{enumerate}

\subsection{Bootloader Updater}
The Bootloader Updater is a specialized component in embedded systems responsible for managing the update process of the bootloader itself. Unlike a standard bootloader, which focuses on system initialization and firmware updates, the Bootloader Updater ensures the bootloader remains up-to-date, secure, and functional. This feature is crucial in systems where the bootloader may require enhancements or patches after deployment.

In our system, two types of Bootloader Updaters are implemented:

\begin{enumerate}
    \item \textbf{Silent Bootloader Updater:} 
    Operates autonomously without user interaction, typically preconfigured to update the bootloader under specific conditions.

    \item \textbf{Communicative Bootloader Updater:} 
    Engages with the host or user for update instructions and provides feedback during the process.
\end{enumerate}

\subsubsection{Bootloader Updater Behavior}
\label{subsec:bootloader_updater_behavior}

The Bootloader Updater is designed to operate seamlessly within the system’s memory structure. Its behavior can be outlined as follows:

\begin{enumerate}
    \item \textbf{System Initialization:} 
    Initializes necessary peripherals, such as communication interfaces and memory controllers, required for update operations.

    \item \textbf{Decision-Making Logic:}
    \begin{itemize}
        \item \textbf{Silent Bootloader Updater:} 
        Automatically decides to proceed with the update based on predefined conditions (e.g., a flag or a specific event).
        \item \textbf{Communicative Bootloader Updater:} 
        Waits for commands from the host or user to initiate the update process.
    \end{itemize}

    \item \textbf{Backup and Recovery:} 
    Before proceeding with the update, the Bootloader Updater creates a backup of the existing bootloader. This ensures the system can recover to a known state if the update fails.

    \item \textbf{Update Process:}
    \begin{itemize}
        \item Receives the new bootloader image via a communication interface.
        \item Verifies the integrity of the received image using CRC or cryptographic techniques.
        \item Programs the new image into the designated memory area.
    \end{itemize}

    \item \textbf{Verification and Handoff:}
    \begin{itemize}
        \item Verifies the integrity of the updated bootloader.
        \item Transfers control to the updated bootloader or returns control to the system, depending on the update mode.
    \end{itemize}
\end{enumerate}

\subsubsection{Bootloader Updater Memory Integration}
\label{subsec:bootloader_updater_memory_integration}

In the system’s memory architecture:

\begin{itemize}
    \item \textbf{Dedicated Memory Segment:} 
    The Bootloader Updater shares the same memory region as the application; specifically at sector 5, ensuring isolation from the primary bootloader memory area.

    \item \textbf{Interaction with Boot Manager:} 
    The Bootloader Updater interacts with the Boot Manager to manage control flow and uses RTC backup registers to store update-related flags.

    \item \textbf{Memory Allocation for Safe Updates:} 
    Separate memory sections are allocated for the existing bootloader, the new bootloader image, and backup data to ensure safe and reliable updates.
\end{itemize}

\subsubsection{Bootloader Updater Main Function}
\begin{enumerate}
    \item \textbf{Silent Bootloader Updater:} \\
    Below is an overview of the silent bootloader updater main function, Update\_Logic():
    \begin{lstlisting}[language=C, caption={Update\_Logic() Function}, label={lst:update_logic}]
void Update_Logic(){
    uint32_t size = sizeof(bootloader_as_array)/sizeof(bootloader_as_array[0]);
    
    // Erase the old bootloader
    Flash_Memory_Erase(BOOTLOADER_BINARY_START_ADDRESS, size);
    
    // Write new bootloader
    Flash_Memory_Write(BOOTLOADER_BINARY_START_ADDRESS, bootloader_as_array, size);

    // Exit handler
    Leaving_Handler();
}
\end{lstlisting}

The \texttt{Update\_Logic()} function operates as follows:

\begin{enumerate}
    \item \textbf{Calculate Bootloader Size:}
    \begin{itemize}
        \item The size of the new Bootloader firmware is determined by dividing the total size of the \texttt{bootloader\_as\_array} by the size of each element in the array.
    \end{itemize}

    \item \textbf{Erase Old Bootloader:}
    \begin{itemize}
        \item The \texttt{Flash\_Memory\_Erase()} function is invoked to clear the memory region starting at \texttt{BOOTLOADER\_BINARY\_START\_ADDRESS}.
        \item The \texttt{size} parameter ensures that the appropriate number of sectors is erased.
    \end{itemize}

    \item \textbf{Write New Bootloader:}
    \begin{itemize}
        \item The \texttt{Flash\_Memory\_Write()} function is called to write the new Bootloader firmware to the cleared memory region.
        \item The function ensures that each byte of the new Bootloader is written sequentially and verifies the data integrity.
    \end{itemize}

    \item \textbf{System Transition:}
    \begin{itemize}
        \item Once the update process is complete, \texttt{Leaving\_Handler()} is called to reset the system and jump to the Boot Manager.
    \end{itemize}
\end{enumerate}

    \item \textbf{Communicative  Bootloader Updater:} \\
    Below is an overview of the Communicative bootloader updater main function, Bootloader\_Updater\_Receive\_Command():
    \begin{lstlisting}[language=C, caption={Bootloader\_Updater\_Receive\_Command() Function}, label={lst:bootloader_updater_receive_command}]
static void Bootloader_Updater_Receive_Command(void) {
    CAN_RxHeaderTypeDef RxHeader;

    // Clear receiving buffer
    memset(Bootloader_Updater_Rx_Buffer, 0, BOOTLOADER_UPDATER_RX_BUFFER_LENGTH);

    // Receive the length of the command via CAN
    HAL_CAN_GetRxMessage(&hcan1, CAN_RX_FIFO0, &RxHeader, Bootloader_Updater_Rx_Buffer);

    // Receive the actual command based on the length
    HAL_CAN_GetRxMessage(&hcan1, CAN_RX_FIFO0, &RxHeader, &Bootloader_Updater_Rx_Buffer[1]);

    switch (Bootloader_Updater_Rx_Buffer[1]) {
        case BOOTLOADER_UPDATER_GET_VERION_COMMAND:
            Get_Version_Command_Handler();
            break;
        case BOOTLOADER_UPDATER_MEM_WRITE_BOOTLOADER_COMMAND:
            Mem_Write_BOOTLOADER_Command_Handler();
            break;
        case BOOTLOADER_UPDATER_MEM_ERASE_BOOTLOADER_COMMAND:
            Mem_Erase_BOOTLOADER_Command_Handler();
            break;
        case BOOTLOADER_UPDATER_LEAVING_TO_BOOT_MANAGER_COMMAND:
            Leaving_To_Boot_Manager_Command_Handler();
            break;
        default:
            // Do nothing for unsupported commands
            break;
    }
}
\end{lstlisting}
The \texttt{Bootloader\_Updater\_Receive\_Command()} function operates as follows:

\begin{enumerate}
    \item \textbf{Buffer Initialization:}
    \begin{itemize}
        \item The \texttt{Bootloader\_Updater\_Rx\_Buffer} is cleared to ensure no residual data interferes with the current command.
    \end{itemize}

    \item \textbf{Receiving the Command Length:}
    \begin{itemize}
        \item The function uses \texttt{HAL\_CAN\_GetRxMessage()} to retrieve the length of the incoming command from the CAN receive FIFO.
    \end{itemize}

    \item \textbf{Receiving the Actual Command:}
    \begin{itemize}
        \item Based on the command length, the function retrieves the command payload, which contains the command code and any associated parameters.
    \end{itemize}

    \item \textbf{Command Parsing:}
    \begin{itemize}
        \item The received command is identified using a \texttt{switch-case} structure.
        \item Each case corresponds to a specific operation, such as retrieving the Bootloader version, writing to memory, or erasing memory.
    \end{itemize}

    \item \textbf{Command Execution:}
    \begin{itemize}
        \item Upon identifying a valid command, the function calls the corresponding handler function:
        \begin{itemize}
            \item \texttt{Get\_Version\_Command\_Handler()}: Retrieves the current Bootloader version.
            \item \texttt{Mem\_Write\_BOOTLOADER\_Command\_Handler()}: Writes data to the Bootloader memory region.
            \item \texttt{Mem\_Erase\_BOOTLOADER\_Command\_Handler()}: Erases the Bootloader memory region.
            \item \texttt{Leaving\_To\_Boot\_Manager\_Command\_Handler()}: Transitions the system back to the Boot Manager.
        \end{itemize}
    \end{itemize}

    \item \textbf{Default Case:}
    \begin{itemize}
        \item For unsupported commands, the function takes no action, ensuring that invalid commands are safely ignored.
    \end{itemize}
\end{enumerate}
\end{enumerate}

\newpage
\subsection{Design Flow}
The system design flow illustrates the seamless interaction of all components, ensuring reliable operation, efficient decision-making, and secure transitions between system states. The following sequence diagram outlines how the system operates:
\begin{figure}[h]
    \centering
    \includegraphics[width=1.0\linewidth]{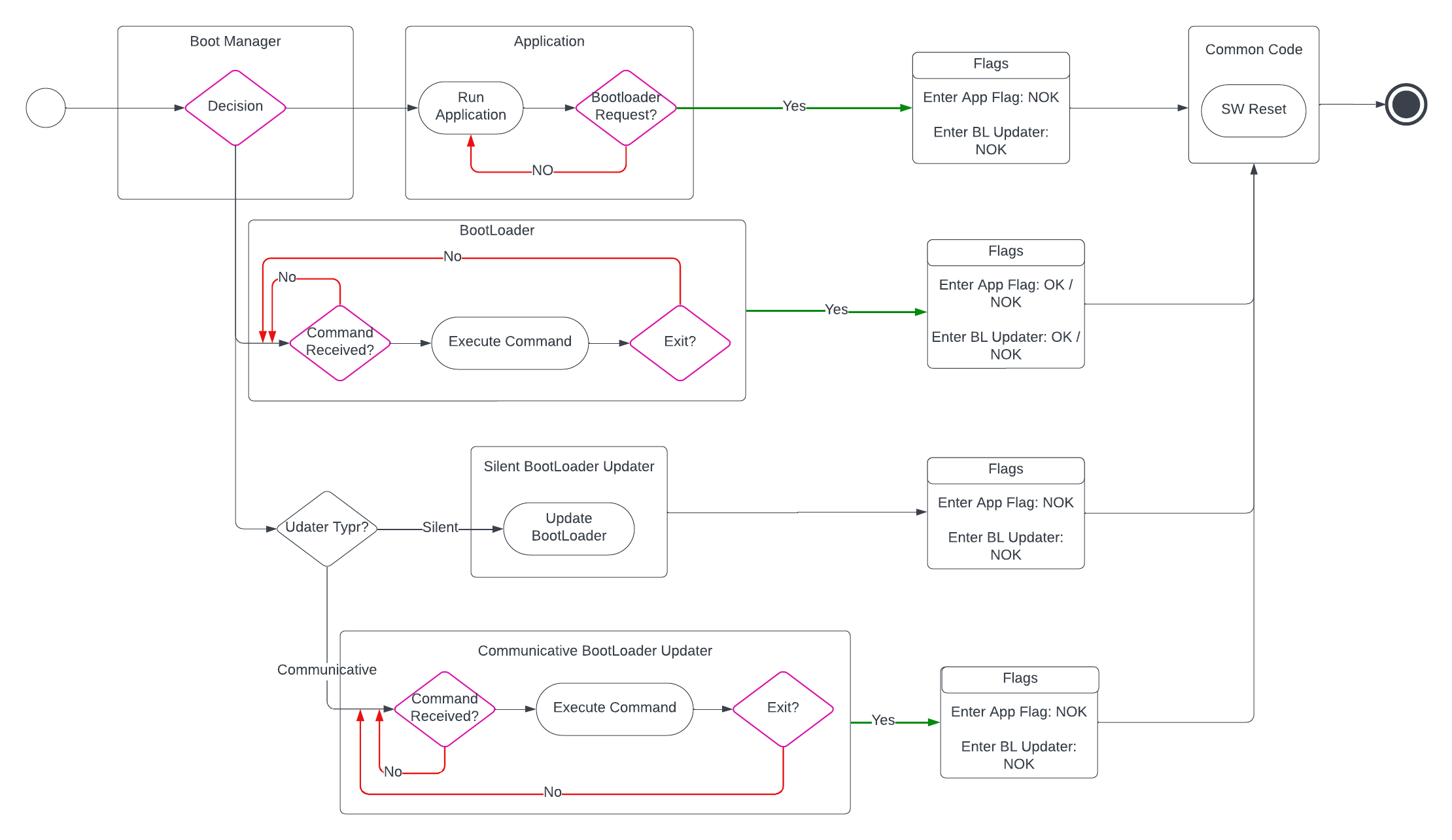}
    \caption{Design Flow}
    \label{fig:enter-label}
\end{figure}
\begin{enumerate}
    \item \textbf{Boot Manager Initialization:}
    \begin{itemize}
        \item The Boot Manager is the first component executed after the system powers on or resets.
        \item It makes a decision based on system flags (\texttt{Enter App Flag} and \texttt{Enter BL Updater Flag}) to determine the next step:
        \begin{itemize}
            \item If the \texttt{Enter App Flag} is set to \texttt{OK}, the system transitions to the Application.
            \item If the \texttt{Enter BL Updater Flag} is set to \texttt{OK}, the system transitions to the Bootloader Updater.
            \item Otherwise, the system defaults to the Bootloader.
        \end{itemize}
    \end{itemize}
    \newpage
    \item \textbf{Application Execution:}
    \begin{itemize}
        \item In the Application, normal operations occur until a Bootloader Request is made.
        \item The Bootloader Request is determined based on application logic or external commands:
        \begin{itemize}
            \item If the request is made, the \texttt{Enter BL Updater Flag} is set to \texttt{OK}, and the system transitions to the Bootloader.
            \item If no request is made, the application continues its execution without interruption.
        \end{itemize}
    \end{itemize}

    \item \textbf{Bootloader Operations:}
    \begin{itemize}
        \item The Bootloader enters a loop to check for incoming commands:
        \begin{itemize}
            \item If a command is received, it is executed (e.g., update, diagnostics, or other tasks).
            \item After executing a command, the bootloader checks if an Exit Condition is met:
            \begin{itemize}
                \item If the exit condition is true, the system transitions to the next state (e.g., Boot Manager or Application).
                \item If the exit condition is not met, the bootloader remains active, waiting for further commands.
            \end{itemize}
        \end{itemize}
    \end{itemize}

    \item \textbf{Transition to Bootloader Updater:}
    \begin{itemize}
        \item If the Boot Manager decides to enter the Bootloader Updater, it determines the type of updater:
        \begin{itemize}
            \item Silent Bootloader Updater
            \item Communicative Bootloader Updater
        \end{itemize}
        \item The type of updater is determined based on a predefined configuration or flag.
    \end{itemize}

    \item \textbf{Silent Bootloader Updater:}
    \begin{itemize}
        \item In the Silent Bootloader Updater, the process is automatic and requires no external interaction.
        \begin{itemize}
            \item The updater directly updates the Bootloader.
            \item After the update is complete, the flags (\texttt{Enter App Flag} and \texttt{Enter BL Updater Flag}) are reset to \texttt{NOK}, and the system performs a Software Reset to transition to the Boot Manager.
        \end{itemize}
    \end{itemize}

    \item \textbf{Communicative Bootloader Updater:}
    \begin{itemize}
        \item The Communicative Bootloader Updater interacts with an external host or controller.
        \begin{itemize}
            \item The updater checks if a command is received from the host.
            \item If a command is received, it executes the specified task (e.g., writing or erasing the bootloader).
            \item After command execution, the updater checks for an Exit Condition:
            \begin{itemize}
                \item If the exit condition is true, the updater resets the flags and performs a Software Reset to transition to the Boot Manager.
                \item If the exit condition is not met, the updater remains active, waiting for further commands.
            \end{itemize}
        \end{itemize}
    \end{itemize}

    \item \textbf{System Flags and Reset:}
    \begin{itemize}
        \item Throughout the system, flags play a crucial role in determining the next state:
        \begin{itemize}
            \item \texttt{Enter App Flag:} Indicates whether to enter the Application.
            \item \texttt{Enter BL Updater Flag:} Indicates whether to enter the Bootloader Updater.
        \end{itemize}
        \item After completing a process (e.g., Bootloader or Updater), the flags are reset to \texttt{NOK}, and a Software Reset is performed to restart the system.
    \end{itemize}
\end{enumerate}

\newpage
\section{CAN Communication Protocol}
\label{subsec:can_implementation}

In this project, the Controller Area Network (CAN) protocol is implemented to facilitate secure and efficient communication between the master ECU and the target ECU during the Firmware Over-the-Air (FOTA) update process. The protocol ensures reliable transfer of the new firmware from the master ECU to the target ECU while adhering to ISO 11898 standards for automotive communication.

\subsection{System Architecture}

The CAN protocol forms the backbone of communication in this project, connecting two primary nodes:

\begin{enumerate}
    \item \textbf{Master ECU:}
    \begin{itemize}
        \item \textbf{Responsibilities:}
        \begin{itemize}
            \item Initiates the FOTA process.
            \item Manages the transfer of firmware data to the target ECU.
            \item Oversees the status of the firmware update process and handles retransmissions if errors are detected.
        \end{itemize}
    \end{itemize}
    \item \textbf{Target ECU:}
    \begin{itemize}
        \item \textbf{Responsibilities:}
        \begin{itemize}
            \item Receives and processes firmware data transmitted by the master ECU.
            \item Validates the integrity of the received data and provides acknowledgment of successful reception.
            \item Programs the validated firmware into the designated Flash memory regions.
        \end{itemize}
    \end{itemize}
\end{enumerate}

\subsection{CAN Protocol Layers}

The implementation of the CAN protocol adheres to the layered architecture defined in ISO 11898, ensuring modularity and clarity in communication design.

\begin{enumerate}
    \item \textbf{Physical Layer:}
    \begin{itemize}
        \item Implements the electrical and hardware specifications required for CAN communication:
        \begin{itemize}
            \item \textbf{Twisted-Pair Cabling:} Used for CANH and CANL to minimize electromagnetic interference (EMI).
            \item \textbf{Differential Signaling:} Ensures robust communication in noisy environments.
            \item \textbf{Termination:} Each end of the CAN bus is terminated with a 120-ohm resistor to maintain proper signal levels.
        \end{itemize}
    \end{itemize}
    \item \textbf{Data Link Layer:}
    \begin{itemize}
        \item Handles data transmission, arbitration, error detection, and acknowledgment:
        \begin{itemize}
            \item \textbf{Arbitration Field:} Enables priority-based message transmission using 11-bit identifiers for standard CAN frames.
            \item \textbf{Error Detection:} Includes mechanisms for detecting bit errors, CRC errors, form errors, and acknowledgment errors.
            \item \textbf{Acknowledgment Field:} Confirms successful message delivery.
        \end{itemize}
    \end{itemize}
    \item \textbf{Application Layer:}
    \begin{itemize}
        \item Implements project-specific logic for FOTA communication:
        \begin{itemize}
            \item \textbf{Command Management:} Defines actions such as starting, pausing, resuming, and ending updates.
            \item \textbf{Data Chunking:} Splits firmware into smaller frames for transmission.
            \item \textbf{Error Recovery:} Ensures data integrity through retransmissions and rollback mechanisms.
        \end{itemize}
    \end{itemize}
\end{enumerate}

\subsection{CAN Frame Usage in FOTA}

The following frame types are used to support the FOTA process:

\begin{enumerate}
    \item \textbf{Command Frames:}
    \begin{itemize}
        \item Control the update process (e.g., start, pause, resume, or end).
        \item Include metadata such as sector numbers or transfer offsets.
    \end{itemize}
    \item \textbf{Data Frames:}
    \begin{itemize}
        \item Carry firmware payloads (up to 8 bytes per frame in standard CAN).
        \item Include sequence numbers for ordered delivery and CRC for data validation.
    \end{itemize}
    \item \textbf{Acknowledgment Frames:}
    \begin{itemize}
        \item Sent by the target ECU to confirm successful reception or indicate errors.
    \end{itemize}
    \item \textbf{Error Frames:}
    \begin{itemize}
        \item Automatically generated by the CAN controller to indicate transmission issues.
    \end{itemize}
\end{enumerate}

\subsection{CAN Communication Workflow}

The FOTA process using the CAN protocol follows a structured workflow to ensure reliable and efficient data transfer:

\begin{enumerate}
    \item \textbf{Initialization:}
    \begin{itemize}
        \item Both ECUs initialize their CAN controllers, setting up key parameters such as:
        \begin{itemize}
            \item \textbf{Baud Rate:} Configured to ensure optimal speed and error-free communication.
            \item \textbf{Operating Mode:} Configured as normal or loopback mode during testing.
            \item \textbf{Message Filters:} Applied to ensure that only relevant frames are processed by each ECU.
        \end{itemize}
        \item The CAN bus is terminated with 120-ohm resistors at both ends to maintain signal integrity and reduce reflections.
    \end{itemize}
    \item \textbf{Command Exchange:}
    \begin{itemize}
        \item The master ECU sends Command Frames to initiate and control the update process. Commands include:
        \begin{itemize}
            \item \textbf{Start Update:} Signals the target ECU to prepare for firmware reception.
            \item \textbf{Pause/Resume Update:} Controls the flow of data in case of interruptions.
            \item \textbf{End Update:} Indicates the completion of data transfer.
        \end{itemize}
    \end{itemize}
    \item \textbf{Data Transmission:}
    \begin{itemize}
        \item Firmware data is divided into small chunks and transmitted as Data Frames. Each frame contains:
        \begin{itemize}
            \item A sequence number to ensure ordered delivery.
            \item A cyclic redundancy check (CRC) value for integrity verification.
        \end{itemize}
        \item The target ECU validates each frame upon reception and stores the data temporarily in a buffer.
    \end{itemize}
    \item \textbf{Acknowledgment and Error Handling:}
    \begin{itemize}
        \item The target ECU sends Acknowledgment Frames after successfully receiving each frame. These frames indicate:
        \begin{itemize}
            \item \textbf{Success:} The frame was received and verified.
            \item \textbf{Error:} An issue occurred during reception, such as CRC mismatch.
        \end{itemize}
        \item In case of errors, the master ECU retransmits the affected frame.
        \item The CAN protocol’s built-in fault confinement mechanisms, such as error-active, error-passive, and bus-off states, ensure network stability even in error scenarios.
    \end{itemize}
    \item \textbf{Firmware Validation and Programming:}
    \begin{itemize}
        \item After all data frames are received, the target ECU validates the firmware using a checksum or hash-based verification.
        \item The validated firmware is then programmed into the Flash memory sectors according to the predefined memory layout.
    \end{itemize}
\end{enumerate}

\newpage
\subsubsection{Transmission Process Flow Chart}
The following flow chart illustrates the transmission process in the CAN communication workflow:
\begin{figure}[h]
    \centering
    \includegraphics[width=0.8\linewidth]{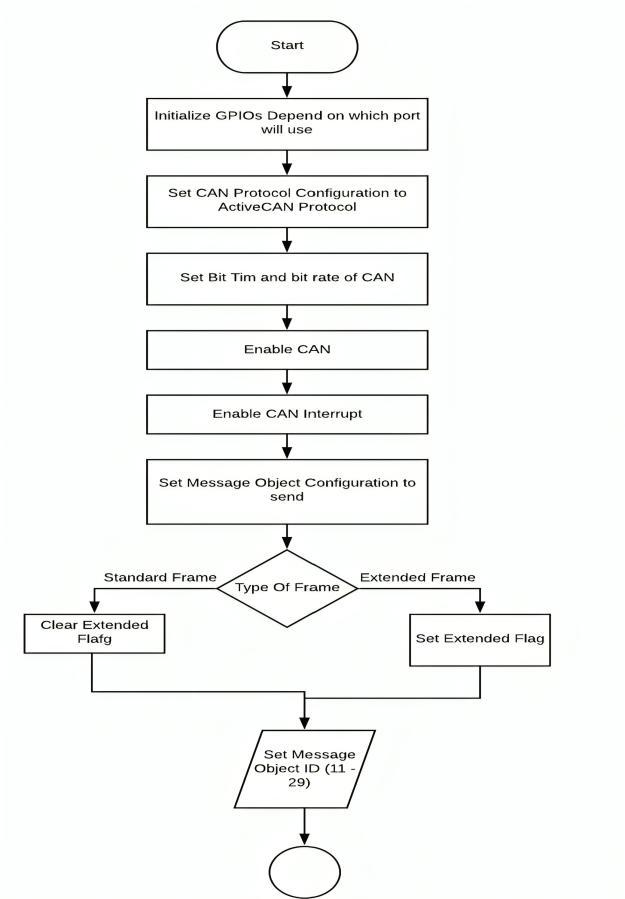}
    \label{fig:transmission-flow-chart}
\end{figure}

\newpage
\begin{figure}[h]
    \centering
    \includegraphics[width=0.75\linewidth]{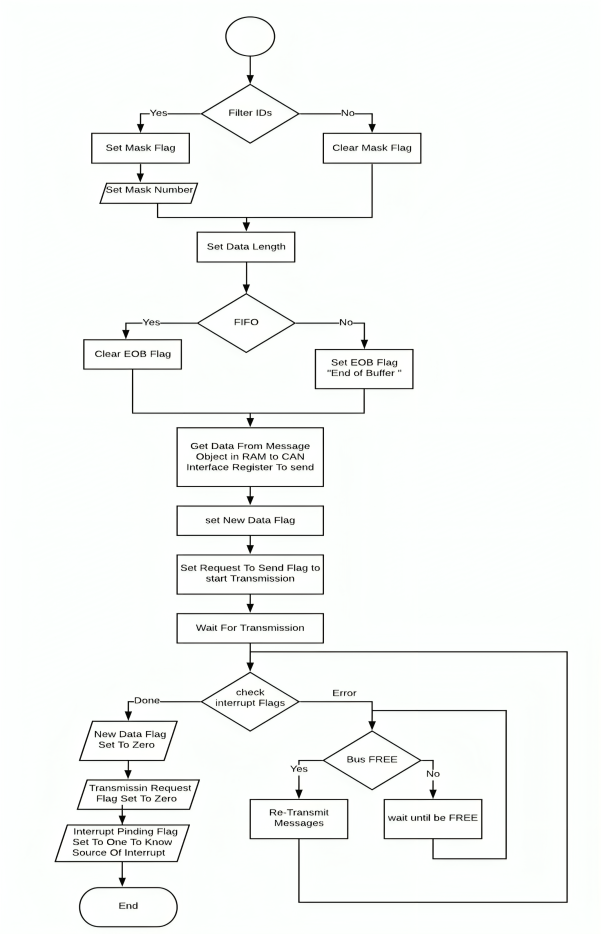}
    \caption{Transmission Process Flow Chart}
    \label{fig:transmission-image1}
\end{figure}

\newpage
\subsubsection{Receiving Process Flow Chart}
The following flow chart illustrates the receiving process in the CAN communication workflow:
\begin{figure}[h]
    \centering
    \includegraphics[width=0.8\linewidth]{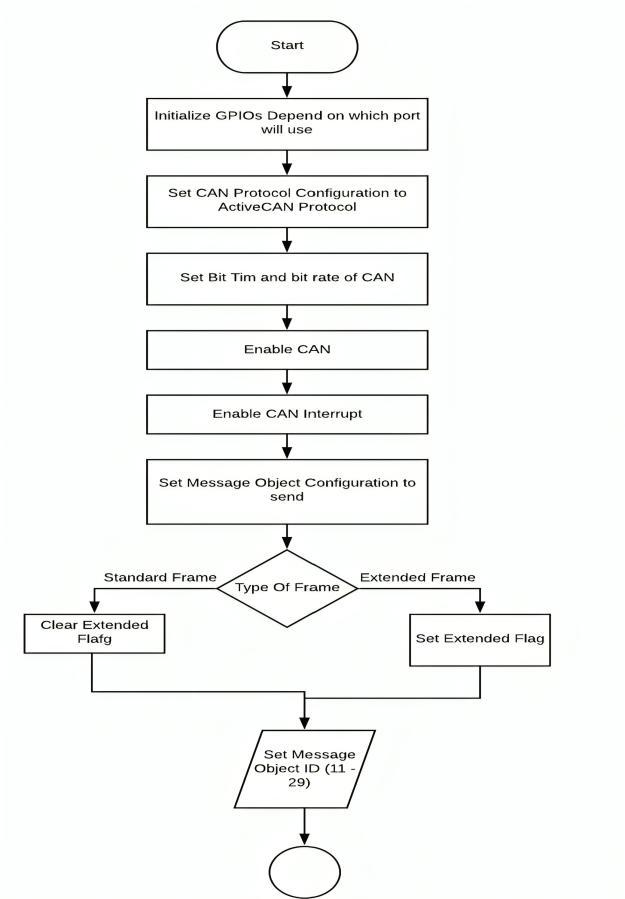}
    \label{fig:receiving-flow-chart}
\end{figure}

\newpage
\begin{figure}[h]
    \centering
    \includegraphics[width=0.75\linewidth]{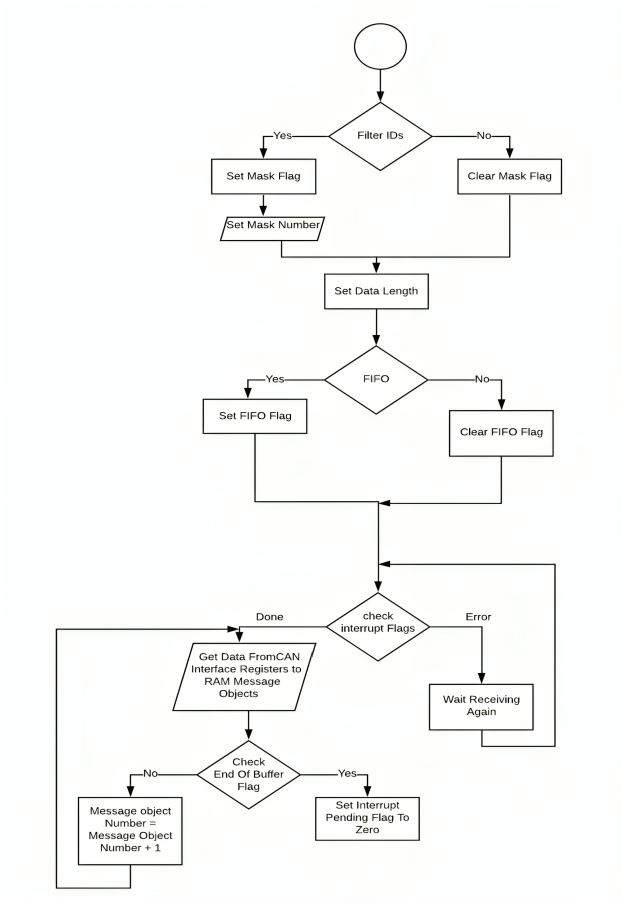}
    \caption{Receiving Process Flow Chart}
    \label{fig:receiving-image1}
\end{figure}

\newpage
\section{Security}
\label{sec:im_security}

\subsection{UDS Protocol: 0x27 Security Access Service}
\label{subsec:uds_0x27_security_access}

In this section, we outline the implementation of the UDS 0x27 Security Access service in our project, which enables secure communication between the diagnostic tester (master) and the Electronic Control Units (ECUs). The service ensures that only authorized entities can interact with sensitive ECU functions, such as firmware updates, calibration, and memory access. 

\subsubsection{Request Frame Format}

The structure of the UDS request frame for Security Access consists of the following elements:

\begin{itemize}
    \item \textbf{Service ID (SID):} The service ID for Security Access is \texttt{0x27}, indicating that the request pertains to the 0x27 Security Access service.
    \item \textbf{Sub-Function Byte (optional):} This byte specifies the action within the Security Access service. It can have two values:
    \begin{itemize}
        \item \texttt{0x01} for "Request Seed"
        \item \texttt{0x02} for "Send Key"
    \end{itemize}
    \item \textbf{Data Parameters:} These are any additional parameters required by the service, such as the seed or key values.
\end{itemize}

\subsubsection{Positive Response Frame}

When the ECU successfully processes the request, it responds with a positive response frame. The structure of the positive response frame is similar to the request frame, but with the first byte (SID) modified as follows:
\begin{itemize}
    \item \textbf{SID + 0x40:} The first byte of the response frame is the SID of the requested service, incremented by \texttt{0x40}. This is used to differentiate between request and response frames.
\end{itemize}

For example, a response to the "Request Seed" frame (\texttt{0x27 0x01}) would have a positive response frame with SID \texttt{0x67}.

\subsubsection{Negative Response Frame}

If the request is invalid or the ECU cannot perform the requested operation, it sends a negative response frame:
\begin{itemize}
    \item \textbf{Negative Response SID (\texttt{0x7F}):} The SID in the negative response frame is always \texttt{0x7F}, indicating an error.
    \item \textbf{Rejected SID:} The requested service ID (\texttt{0x27} in this case) is included in the negative response.
    \item \textbf{NRC (Negative Response Code):} This byte provides details about the error, such as:
    \begin{itemize}
        \item \texttt{0x22} - Conditions Not Correct
        \item \texttt{0x24} - Request Sequence Error
        \item \texttt{0x35} - Invalid Key
        \item \texttt{0x36} - Exceeded Number of Attempts
    \end{itemize}

\end{itemize}

\subsubsection{Security Access Process}

The Security Access service requires a two-step authentication process: requesting a seed and sending the unlock key. The communication process is as follows:

\begin{enumerate}
    \item \textbf{Request Seed:}
    \begin{itemize}
        \item The master (diagnostic tester) sends a request to the target ECU to initiate the security access process. The request includes SID \texttt{0x27} and Sub-Function \texttt{0x01}.
        \item The ECU generates a random seed using a random number generator peripheral and sends it back to the master with a positive response. The response frame includes SID \texttt{0x67}, Sub-Function \texttt{0x01}, and the generated seed.
    \end{itemize}
    \item \textbf{Send Key:}
    \begin{itemize}
        \item Using the received seed, the master calculates the corresponding unlock key and sends it to the target ECU. The request frame includes SID \texttt{0x27} and Sub-Function \texttt{0x02}, followed by the 32-byte key.
        \item If the key is correct, the ECU grants access by sending a positive response with SID \texttt{0x67} and Sub-Function \texttt{0x02}. If the key is invalid, the ECU responds with a negative response (\texttt{0x7F}), specifying the error with the corresponding NRC (e.g., \texttt{0x35} for invalid key).
    \end{itemize}
\end{enumerate}

\subsubsection{Security Access Example Frame}

\begin{itemize}
    \item \textbf{Request Seed Frame:}
    \begin{itemize}
        \item SID: \texttt{0x27}
        \item Sub-Function: \texttt{0x01}
        \item (No additional data required)
    \end{itemize}
    \item \textbf{Positive Response Frame (with Seed):}
    \begin{itemize}
        \item SID: \texttt{0x67}
        \item Sub-Function: \texttt{0x01}
        \item Seed: \texttt{xx xx xx xx ...} (Generated by the ECU)
    \end{itemize}
    \item \textbf{Send Key Frame:}
    \begin{itemize}
        \item SID: \texttt{0x27}
        \item Sub-Function: \texttt{0x02}
        \item Key: \texttt{yy yy yy yy ...} (Generated by the master using the seed)
    \end{itemize}
    \item \textbf{Positive Response (Access Granted):}
    \begin{itemize}
        \item SID: \texttt{0x67}
        \item Sub-Function: \texttt{0x02}
    \end{itemize}
    \item \textbf{Negative Response (Invalid Key):}
    \begin{itemize}
        \item SID: \texttt{0x7F}
        \item Rejected SID: \texttt{0x27}
        \item NRC: \texttt{0x35} (Invalid Key)
    \end{itemize}
\end{itemize}

\subsubsection{Sequence of Operation}

\begin{enumerate}
    \item The master requests a seed from the ECU by sending a frame with SID \texttt{0x27} and Sub-Function \texttt{0x01}.
    \item The ECU responds with the generated seed (positive response).
    \item The master computes the key from the seed and sends it to the ECU in a frame with SID \texttt{0x27} and Sub-Function \texttt{0x02}.
    \item The ECU either grants access (positive response) or denies it (negative response with NRC code, such as \texttt{0x35} for invalid key).
\end{enumerate}

By implementing this process, we ensure secure access to sensitive ECU functions, enhancing the integrity and safety of the firmware update process. This approach is aligned with the ISO 14229 standard and ensures that firmware updates are only performed by authorized entities.

\newpage

\section{Memory Stack and Delta Update}

\subsection{Understanding STM32 Flash Characteristics}
\label{subsec:Understanding STM32 Flash Characteristics}

The STM32F401RE microcontroller features internal Flash memory divided into sectors of varying sizes. For this project, the Flash memory was partitioned into \textbf{80 logical NVM blocks}, each 1 KB in size, to meet the requirements of \textbf{delta updates}. This block size was carefully chosen to ensure efficient memory utilization and to optimize the update process by minimizing unnecessary write and erase operations.

The STM32 Flash controller requires a specific unlock and lock sequence for executing write and erase operations. These characteristics dictated the modifications needed in the AUTOSAR Flash driver to ensure compatibility with the STM32 Flash interface.
\subsection{Adapting the AUTOSAR Flash Driver (Fls)}
\label{subsec:Adapting the AUTOSAR Flash Driver (Fls)}

The AUTOSAR Flash driver (Fls) was customized to comply with STM32F401RE Flash specifications. Key modifications include:

\begin{enumerate}
    \item \textbf{Unlocking the Flash Controller}

The unlock sequence for the Flash controller was implemented to allow write and erase operations. This involved writing specific unlock keys to the Flash control registers.
    \item \textbf{Sector Erase Implementation}
    \begin{itemize}
        \item The erase functionality was adapted to use STM32's sector-based erase commands.
        \item The Flash controller's \textbf{Busy (BSY)} flag was monitored to ensure completion of the erase process.
    \end{itemize}
    \item \textbf{Write Function Implementation}
    \begin{itemize}
        \item Data was written in \textbf{32-bit words}, adhering to STM32 Flash write alignment requirements.
        \item A read-back verification mechanism was added to confirm data integrity after writing.
    \end{itemize}
    \item \textbf{Read Function Implementation}
The read functionality was implemented to ensure accurate data retrieval from specific memory locations while adhering to alignment and boundary requirements.

\end{enumerate}
\subsection{Configuration of the Memory Stack}
\label{subsec:Configuration of the Memory Stack}

\begin{enumerate}
    \item \textbf{Logical Block Configuration}
    \begin{itemize}
        \item The Flash memory was divided into 128 logical blocks, each sized at \textbf{1 KB}, to accommodate delta update operations.
        \item Block IDs and attributes, such as CRC validation and redundancy, were configured to enhance reliability and data integrity.
    \end{itemize}
    \item \textbf{NvM Module Customization}
    \begin{itemize}
        \item The NvM module was configured to manage these logical blocks, enabling the application to store, retrieve, and update data efficiently.
        \item Specific attributes, such as error detection via CRC and block management strategies, were implemented to meet the project’s robustness requirements.
    \end{itemize}
\end{enumerate}
\subsection{Delta Update Implementation Using CRC}
\label{subsec:Delta Update Implementation Using CRC}

Delta updates were implemented to minimize memory usage by storing only the modified portions of the firmware image. A \textbf{Cyclic Redundancy Check (CRC)} mechanism was utilized to ensure data integrity during the update process. The steps for delta update implementation are as follows:

\begin{enumerate}
    \item \textbf{Initial CRC Calculation}
    \begin{itemize}
        \item Before the update, the current firmware in Flash memory is divided into logical blocks.
        \item The CRC of each block is calculated and stored in a designated metadata section.
    \end{itemize}
    \item \textbf{Comparison During Update}
    \begin{itemize}
        \item During a delta update, the incoming firmware is divided into the same logical blocks.
        \item The CRC of each incoming block is calculated and compared to the CRC of the corresponding block in Flash memory.
    \end{itemize}
    \item \textbf{Selective Erase and Write}
    \begin{itemize}
        \item If the CRC of an incoming block differs from the stored CRC, the corresponding block in Flash memory is erased and updated with the new data.
        \item Blocks with matching CRCs are skipped, reducing unnecessary write and erase cycles.
    \end{itemize}
    \item \textbf{Final Verification}
After the update, the CRCs of the updated blocks are recalculated and compared with the incoming firmware to ensure successful updates.

\end{enumerate}

\subsection{Sequence of the Delta Update Process}
\label{subsec:Sequence of the Delta Update Process}

The sequence of operations for delta updates, integrated with the NVM main function and FreeRTOS, is as follows:

\begin{enumerate}
    \item \textbf{Triggering the Update}
    \begin{itemize}
        \item The system periodically checks for new firmware updates every 15 minutes using a FreeRTOS task.
        \item If a new firmware image is detected, the update process is initiated.
    \end{itemize}
    \item \textbf{Data Reception and CRC Comparison}
    \begin{itemize}
        \item The firmware image is received in 1 KB chunks.
        \item For each chunk, the CRC is calculated and compared with the stored CRC for the corresponding block in Flash memory.
    \end{itemize}
    \item \textbf{Erase and Write Operations}
If the CRCs differ, the NVM main function performs the following tasks:

    \begin{itemize}
        \item Issues a request to erase the target block.
        \item Polls the Flash controller’s BSY flag to confirm the erase operation is complete.
        \item Writes the new data into the block using the 32-bit word write sequence.
        \item Reads back the written data to verify its integrity.
    \end{itemize}

    \item \textbf{Update Metadata}
After a block is successfully updated, its new CRC is calculated and stored in the metadata section.

    \item \textbf{Post-Update Validation}
    \begin{itemize}
        \item Once all blocks are processed, the firmware integrity is validated by recalculating and verifying the CRCs of all blocks.
        \item If the validation succeeds, the system marks the update as complete and reboots into the updated firmware.
    \end{itemize}
\end{enumerate}

\subsection{Integration of the Flash Driver into the AUTOSAR Stack}
\label{subsec:Integration of the Flash Driver into the AUTOSAR Stack}

The customized Flash driver (Fls) was integrated with the Memory Abstraction Layer (MemIf) to ensure modularity and seamless communication with the higher-level NvM module. The integration enables the following key operations:

\begin{itemize}
    \item \textbf{Read}: Retrieves data from a specified memory address.
    \item \textbf{Write}: Programs data into the Flash memory in 32-bit words.
    \item \textbf{Erase}: Clears specific sectors of Flash memory.
\end{itemize}
The \textbf{NvM\_MainFunction} executes periodically as a FreeRTOS task to handle queued read, write, and erase requests. This ensures real-time operation without disrupting the main application’s workflow.

\subsection{Testing and Validation}
\label{subsec:Testing and Validation}

\begin{enumerate}
    \item \textbf{Write and Verify}

Data was written to specific memory blocks, and read-back verification confirmed data integrity.
    \item \textbf{Erase and Update}

Sectors were erased and updated with new firmware blocks, ensuring proper functionality of the erase and write operations.
    \item \textbf{Delta Update Simulation}
    \begin{itemize}
        \item A simulated delta update was performed by introducing changes to specific firmware blocks.
        \item Only modified blocks were updated, demonstrating the efficiency of the delta update mechanism.
    \end{itemize}
    \item \textbf{Boundary and Error Conditions}
    \begin{itemize}
        \item Boundary cases, such as overlapping blocks and partial updates, were tested to ensure robustness.
        \item Error scenarios, including failed CRC validation, were simulated to verify proper error handling.
    \end{itemize}
\end{enumerate}

\newpage

\section{Lane Keeping Assist (LKA) Implementation}

The Lane Keeping Assist (LKA) system is designed to autonomously control the vehicle's steering by integrating a Python-based vision processing module with an STM32F401RE microcontroller for motor control. This section details the main functions implemented in both the Python script and the STM32 firmware.
\subsection{Python Lane Detection Functions}

The Python script processes video input, detects lanes, calculates deviations, and communicates with the STM32 microcontroller. Below are the key functions \cite{lane_detection_steer_departure}:

\begin{lstlisting}[language=Python]
def offCenter(meanPts, inpFrame): 
\end{lstlisting}

    \begin{itemize}
        \item \textbf{Purpose:}  
        Calculates the vehicle's lateral deviation from the lane center and determines the appropriate steering direction.

        \item \textbf{Inputs:}  
        \begin{itemize}
            \item \texttt{meanPts}: The averaged lane center points from the detected lane lines.
            \item \texttt{inpFrame}: The input frame from the camera used for lane detection.
        \end{itemize}

        \item \textbf{Outputs:}  
        \begin{itemize}
            \item \texttt{deviation}: Lateral deviation from the lane center in meters.
            \item \texttt{direction}: Indicates the direction of the deviation (\texttt{``left''} or \texttt{``right''}).
            \item \texttt{Motor\_Order}: Command for the motor:
            \begin{itemize}
                \item \texttt{1}: Turn right.
                \item \texttt{2}: Turn left.
                \item \texttt{3}: Move straight.
            \end{itemize}
        \end{itemize}

        \item \textbf{Functionality:}  
        Calculates the pixel-based deviation by determining the horizontal difference between the image center and the detected lane center. Converts pixel deviation into a physical deviation in meters using a predefined conversion factor (\texttt{xm\_per\_pix}) and determines the corresponding motor command.
    \end{itemize}
\newpage
\begin{lstlisting}[language=Python]
def measure_lane_curvature(ploty, leftx, rightx): 
\end{lstlisting}
    \begin{itemize}
        \item \textbf{Purpose:}  
        Measures the lane curvature in meters and determines the curve's direction (\texttt{``Left Curve''}, \texttt{``Right Curve''}, or \texttt{``Straight''}).

        \item \textbf{Inputs:}  
        \begin{itemize}
            \item \texttt{ploty}: Y-coordinates of the lane points in pixel space.
            \item \texttt{leftx}, \texttt{rightx}: X-coordinates of the left and right lane lines in pixel space.
        \end{itemize}

        \item \textbf{Outputs:}  
        \begin{itemize}
            \item \texttt{curvature}: Average curvature of the lane in meters.
            \item \texttt{curve\_direction}: Indicates the curve direction.
        \end{itemize}

        \item \textbf{Functionality:}  
        Fits a second-degree polynomial to the detected lane lines in world coordinates, calculates the radius of curvature, and determines the curve's direction based on the lane lines' relative positions.
    \end{itemize}

\begin{lstlisting}[language=Python]
def send_float(ser, float_value): 
\end{lstlisting}
    \begin{itemize}
        \item \textbf{Purpose:}  
        Sends a floating-point value (e.g., deviation or curvature) from the Python script to the STM32 microcontroller via serial communication.

        \item \textbf{Inputs:}  
        \begin{itemize}
            \item \texttt{ser}: Serial communication object for interfacing with the STM32.
            \item \texttt{float\_value}: Floating-point value to send.
        \end{itemize}

        \item \textbf{Functionality:}  
        Formats the floating-point value to a string with two decimal places and sends it as bytes over the serial interface. Prints the sent data for debugging purposes.
    \end{itemize}

\newpage
\subsection{STM32 Motor Control Functions}

The STM32 microcontroller receives deviation data from the Python script and adjusts the steering motor to minimize the deviation. Below are the key functions in the STM32 firmware:

\begin{lstlisting}[language=C]
   void MotorControl(float deviation, int16_t encoder_position);
\end{lstlisting}
  
    \begin{itemize}
        \item \textbf{Purpose:}  
        Controls the steering motor to correct the vehicle's deviation and maintain lane centering.

        \item \textbf{Inputs:}  
        \begin{itemize}
            \item \texttt{deviation}: Lateral deviation from the lane center (in meters).
            \item \texttt{encoder\_position}: Current position of the steering motor, measured by the encoder.
        \end{itemize}

        \item \textbf{Functionality:}  
        Computes the PID control output using the deviation and the motor's current position. Adjusts the target encoder position based on the magnitude and direction of the deviation. Controls the motor's speed and direction by updating the PWM signal, ensuring that the motor aligns the vehicle with the lane center. Stops the motor when the target position is reached.
    \end{itemize}

\subsection{System Workflow}

\begin{enumerate}
    \item The Python script processes video input to detect lane boundaries, calculates the deviation from the lane center, and sends the deviation value to the STM32 microcontroller.
    \item The STM32 microcontroller receives the deviation data and calculates the necessary PID adjustments for the motor.
    \item Based on the deviation and motor encoder feedback, the STM32 adjusts the motor's direction and speed to correct the steering.
    \item The vehicle's steering mechanism is continuously adjusted to maintain lane centering.
\end{enumerate}

\section{Steering Wheel System}
\label{sec:steering}

\subsection{Steering System Overview}
\label{subsec:steering_overview}
The steering system plays a fundamental role in implementing autonomous lane-keeping functionality. It is designed to provide precise control over the vehicle's steering angle, ensuring stable lane-centering adjustments. The system integrates multiple mechanical and electrical components that work together to achieve the required performance.

The mechanical structure consists of a steering wheel, a DC motor, and a power transmission system using timing pulleys and belts to transfer motion efficiently. The sensory component includes an optical encoder, which provides real-time feedback on the steering angle and velocity, enabling precise control. The electrical system comprises the Cytron motor driver and the STM32F401RE Nucleo microcontroller, which process encoder data and generate control signals to adjust the motor’s operation. \cite{SharedAutonomySteering2024}.

By combining these elements, the steering system ensures smooth, reliable, and accurate lane-keeping control, forming a crucial part of the overall autonomous driving framework. In the following sections, we will explore each component in detail, highlighting its function and significance in the system.

\begin{figure}[H]
    \centering
    \begin{subfigure}[b]{0.45\linewidth}
        \includegraphics[width=0.75\linewidth]{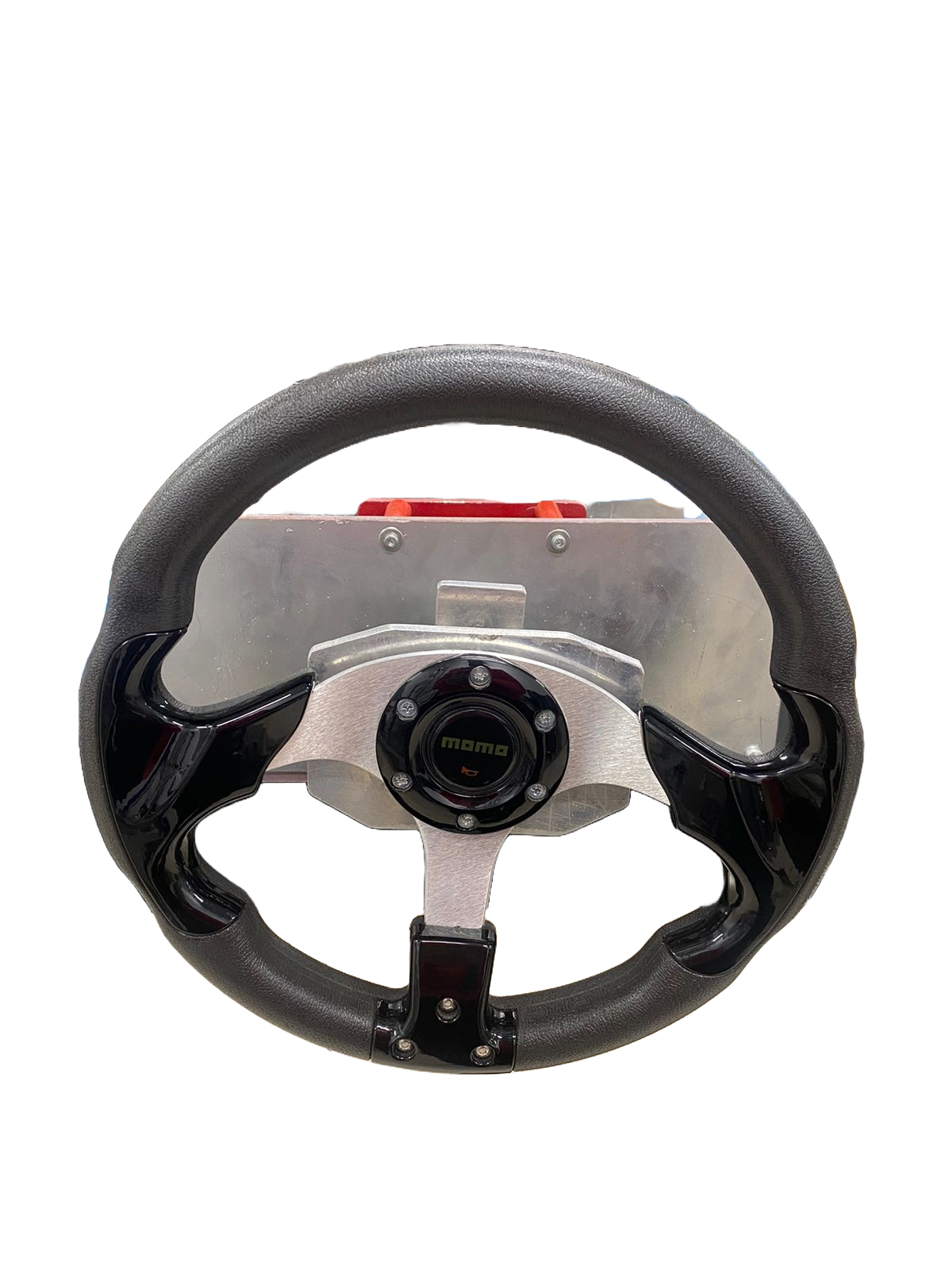}
    \end{subfigure}
    \quad 
    \begin{subfigure}[b]{0.45\linewidth}
        \includegraphics[width=0.75\linewidth]{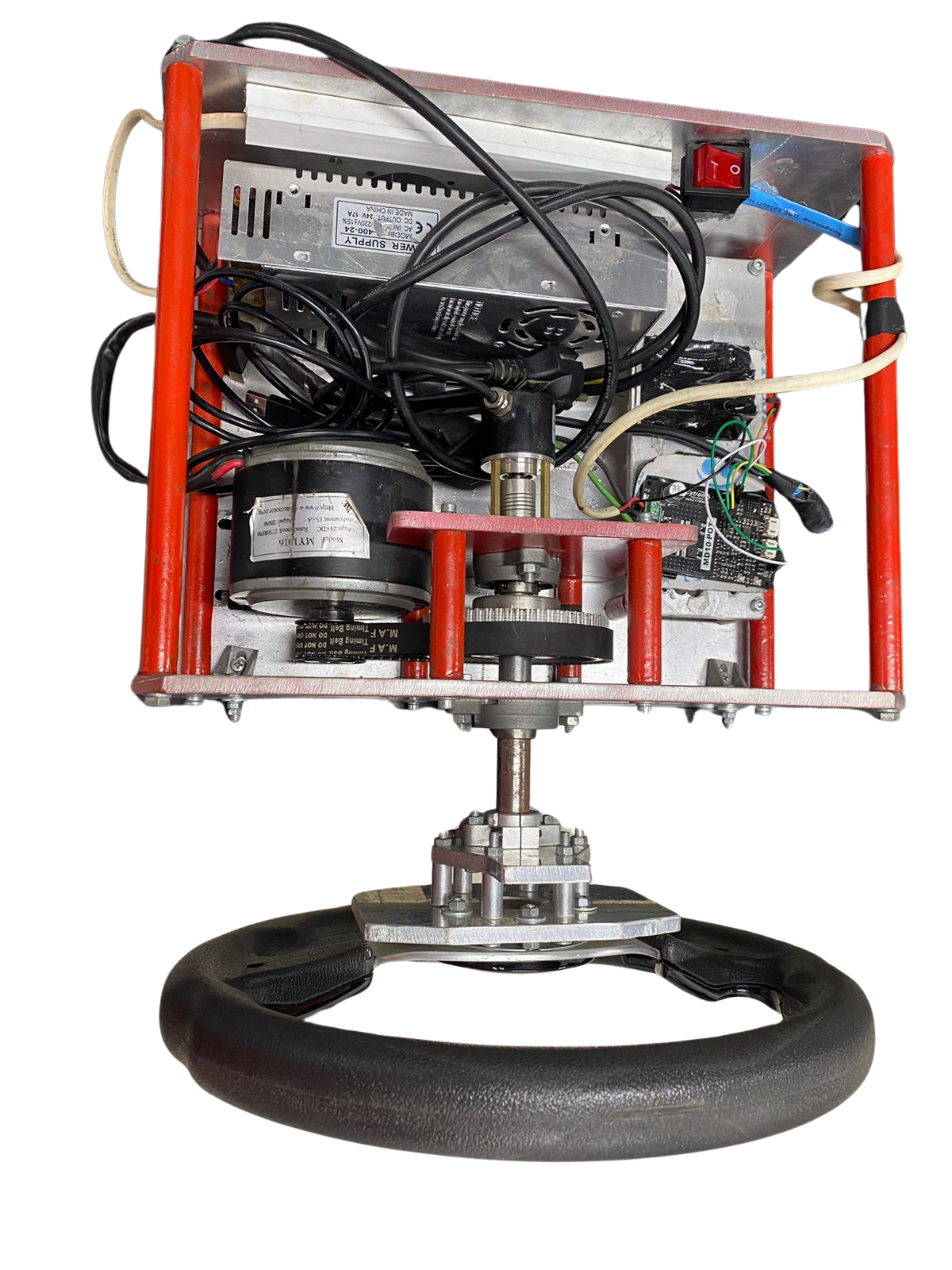}
    \end{subfigure}
    \vspace{1em} 
    \caption{Steering system prototype \cite{SharedAutonomySteering2024}}
    \label{fig:ds1}
\end{figure}
\newpage
\subsubsection{Sensory Component: Optical Encoder}
The optical encoder used in the steering system provides precise angle and angular velocity measurements, crucial for maintaining lane-centering control. The encoder offers a wide operating voltage range (5 to 24 VDC) and boasts high resolution with 2000 pulses per revolution (PPR), which significantly improves measurement accuracy. This encoder also features a zero index (phase Z) for easy adjustment and precise positioning. With its rugged construction, the encoder can withstand both radial and thrust loads, ensuring consistent performance. The choice of an incremental encoder is due to its simplicity and effectiveness in detecting changes in steering position and velocity.

\subsubsection{Actuation Component: DC Brushed Motor}
The DC brushed motor is a key actuator in the steering system, responsible for providing the necessary torque to achieve lane-centering functionality. The required steering torque during normal driving is typically between 0 and 2 N·m. To ensure sufficient torque for autonomous lane-keeping control, a motor capable of providing approximately 3 N·m is used. The selected AlveyTech 24V 250W MY1016 DC motor operates at speeds of 2600-2850 RPM, delivering a minimum torque of around 1 N·m. To fine-tune the motor's output and achieve precise control, a pulley and belt mechanism with a speed reduction ratio of 3:1 is employed.

\begin{figure}[h]
    \centering
    \includegraphics[width=5cm]{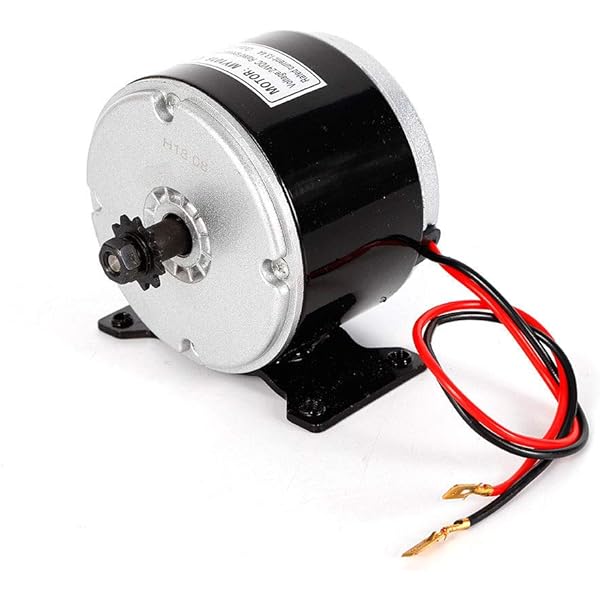}
    \caption{AlveyTech DC Motor \cite{AlveyTech_DC_Motor}}
    \label{fig:dcMotor}
\end{figure}
\newpage
\subsubsection{Electrical Circuit Design}
The electrical system is responsible for controlling the motor and regulating its speed and direction. The Cytron MD10-POT motor driver plays a crucial role in this by providing bi-directional control for the 24V DC motor. This driver uses high-resolution PWM (Pulse Width Modulation) control to achieve precise steering adjustments. The driver is integrated with the STM32F401RE Nucleo board, which processes encoder data and generates the necessary control signals to enable autonomous lane-centering.
\begin{figure}[H]
    \centering
    \includegraphics[width=0.50\linewidth]{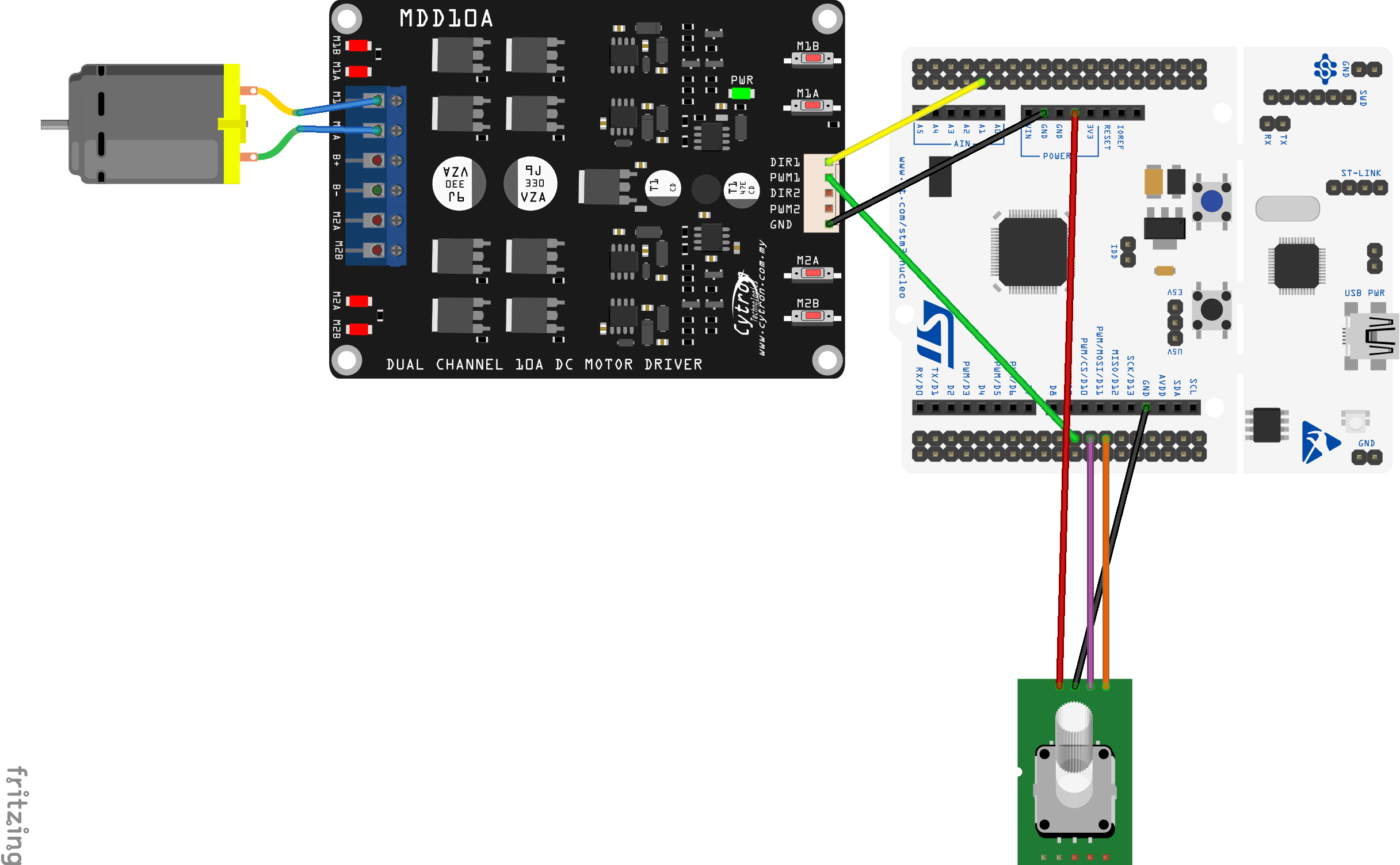}
    \caption{Steering System Schematic}
    \label{fig:enter-label}
\end{figure}
\subsubsection{Electrical Components}
\paragraph{Cytron Motor Driver}\mbox{}\\
The Cytron MD10-POT motor driver is a robust and reliable component that allows for fine-grained control over the motor's direction and speed. It supports a maximum continuous current of 10A and peak currents of up to 30A, making it suitable for controlling the high-power DC motor used in the steering system. The driver accepts both 3.3V and 5V logic-level inputs and operates at PWM frequencies up to 10kHz.

\begin{figure}[H] 
    \centering 
    \includegraphics[width=0.5\linewidth]{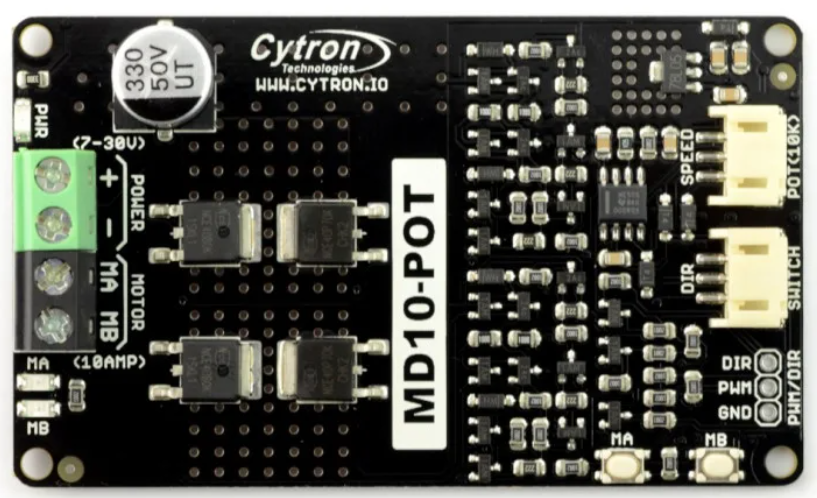} 
    \caption{Cytron MD10-POT Motor Driver, \cite{Cytron_Motor_Driver}} 
    \label{fig:cytron-motor-driver} 
\end{figure}

\paragraph{Microcontroller: STM32F401RE Nucleo Board}\mbox{}\\
The STM32F401RE Nucleo board serves as the central processing unit for the steering system, handling the processing of encoder data and generating the motor control signals. It communicates with higher-level control algorithms through interfaces like UART and enables real-time steering adjustments for lane-keeping applications.

\begin{figure}[H] 
    \centering 
    \includegraphics[width=0.35\linewidth]{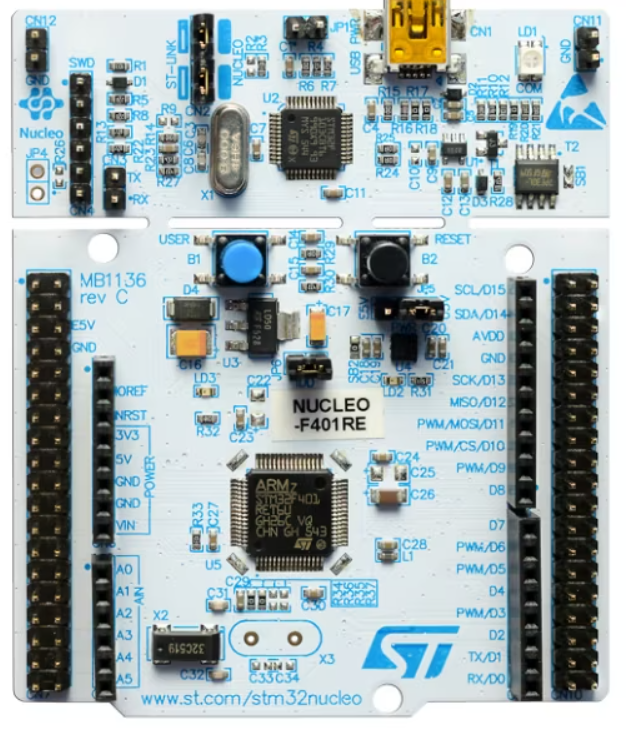} 
    \caption{STM32F401RE Nucleo Board\cite{STM32F401RE_Nucleo}} 
    \label{fig:stm32f401re} 
\end{figure}

\paragraph{Power Supply}\mbox{}\\
A 24V, 17A DC power supply is used to meet the power demands of the motor and control system, ensuring stable and continuous operation during testing and in real-world applications.

\begin{figure}[ht] 
    \centering 
    \includegraphics[width=0.5\linewidth]{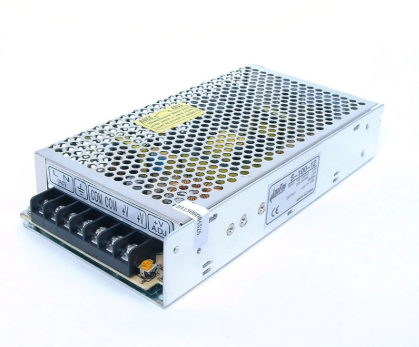} 
    \caption{AmpFlow S-400-24 Power Supply \cite{AmpFlow_Power_Supply}} 
    \label{fig:power-supply} 
\end{figure}

\clearemptydoublepage 

\chapter{Results and Discussion}
\label{chap:ERD}

\section{Integration Testing Results}
Integration tests were conducted to verify the seamless communication and interaction between different system components, including the \textbf{Firmware Over-the-Air (FOTA) update mechanism}, the \textbf{Unified Diagnostic Services (UDS) security authentication}, and the \textbf{Lane Keep Assist (LKA) steering system}. The primary objectives were to measure the firmware flashing time, ensure correct UDS authentication, and validate real-time motor control for LKA.

\subsection{Firmware Flashing Time Analysis}
To evaluate the efficiency of \textbf{delta updating}, we measured the time required to flash the firmware onto the Electronic Control Unit (ECU) before and after implementing the \textbf{delta update technique}. The results are summarized in \textbf{Table~\ref{tab:flashing_time}}:

\begin{table}[h]
    \centering
    \begin{tabular}{|c|c|c|}
        \hline
        \textbf{Update Method} & \textbf{Flashing Time (mm:ss)} & \textbf{Reduction (\%)} \\
        \hline
        Full Firmware Update & 2:35 & - \\
        Delta Firmware Update & 1:15 & 51.6\% Reduction \\
        \hline
    \end{tabular}
    \caption{Firmware Flashing Time Comparison}
    \label{tab:flashing_time}
\end{table}

As shown, the flashing time \textbf{decreased by approximately 51.6\%}, demonstrating the efficiency of \textbf{delta updating} in reducing update duration and improving vehicle availability.

\subsection{UDS Security Authentication Testing}
The UDS \textbf{0x27 (Security Access Service)} was tested to validate its role in ensuring secure firmware updates. The following aspects were analyzed:

\begin{itemize}
    \item \textbf{Challenge-Response Authentication Time:} The time taken to complete the authentication handshake was measured using a stopwatch.
    \item \textbf{Successful Authentication Rate:} The percentage of successful authentications over multiple test iterations.
\end{itemize}

\begin{table}[h]
    \centering
    \begin{tabular}{|c|c|}
        \hline
        \textbf{Test Parameter} & \textbf{Result} \\
        \hline
        Authentication Time (avg.) & 4.2 seconds \\
        Successful Authentication Rate & 100\% \\
        \hline
    \end{tabular}
    \caption{UDS Security Authentication Test Results}
    \label{tab:uds_auth}
\end{table}

The results confirm that the \textbf{UDS security access service is functioning correctly}, allowing only authorized firmware updates.

\section{Validation Testing Results}
Validation tests were conducted to evaluate the \textbf{real-world performance of the Lane Keep Assist (LKA) system}, focusing on \textbf{steering response accuracy and motor control stability}. The \textbf{LKA system used pre-recorded videos of a car moving in streets with lane markings}, with lane deviation measured using \textbf{OpenCV in a Python script}. The deviation values were transmitted to the \textbf{STM32F401RE Nucleo board via USART2}, which controlled the \textbf{DC motor connected to the steering wheel through a Cytron motor driver and encoder feedback}.
\begin{figure}[h]
    \centering
    \includegraphics[width=0.75\linewidth]{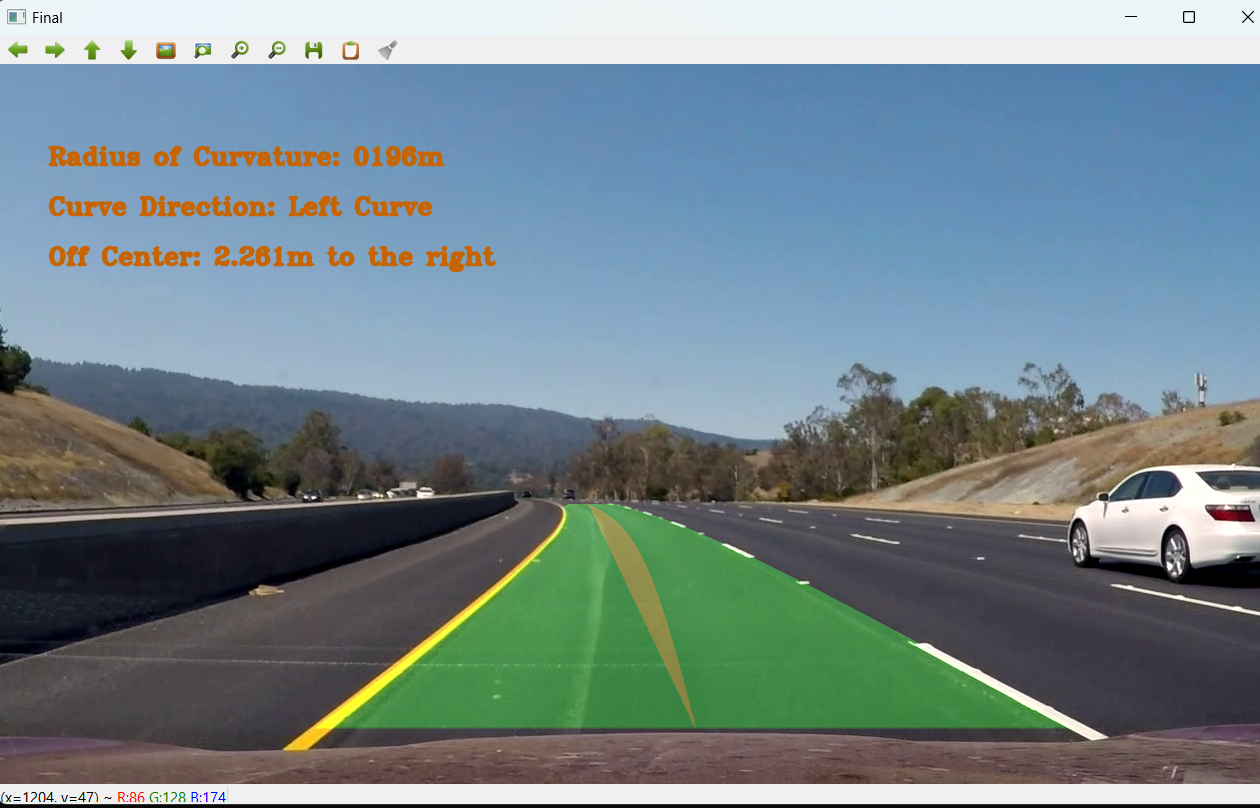}
    \caption{Lane detection results from a camera-mounted car on a highway.}
    \label{fig:enter-label}
\end{figure}

\subsection{Error Reduction Over Time}
The LKA system was tested by setting various target positions corresponding to zero lane deviation. The \textbf{PID controller tuning was adjusted} to minimize the error between the current position and the target position. The system was evaluated based on \textbf{how effectively the error decreases over time}. The results are summarized in \textbf{Table~\ref{tab:steering_correction}}:

\begin{table}[H]
    \centering
    \begin{tabular}{|c|c|c|}
        \hline
        \textbf{Target Position (°)} & \textbf{Initial Error (°)} & \textbf{Final Error (°)} \\
        \hline
        10° & 6° & 0.5° \\
        20° & 12° & 0.7° \\
        30° & 18° & 1.0° \\
        \hline
    \end{tabular}
    \caption{Steering Position Error Reduction}
    \label{tab:steering_correction}
\end{table}

\begin{figure}[h]
    \centering
    \begin{tikzpicture}
        \begin{axis}[
            xlabel={Time (s)},
            ylabel={Steering Position Error (°)},
            legend pos=north east,
            grid=major,
            width=0.8\textwidth,
            height=0.5\textwidth
        ]
        \addplot[color=red, thick] coordinates {
            (0, 10) (0.5, 8) (1, 6) (1.5, 5) (2, 4.5) (2.5, 4.2) (3, 4) (3.5, 3.8) (4, 3.7) (4.5, 3.6) (5, 3.5)
        };
        \addlegendentry{Before Tuning}
        
        \addplot[color=blue, thick] coordinates {
            (0, 10) (0.5, 6) (1, 3.5) (1.5, 2) (2, 1) (2.5, 0.6) (3, 0.4) (3.5, 0.3) (4, 0.2) (4.5, 0.1) (5, 0)
        };
        \addlegendentry{After Tuning}
        \end{axis}
    \end{tikzpicture}
    \caption{PID Response Before and After Tuning}
    \label{fig:pid_response}
\end{figure}
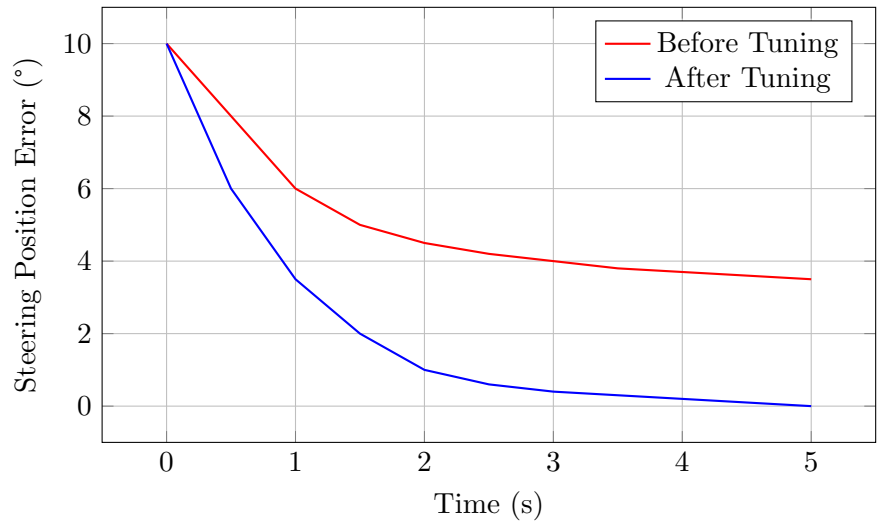

The results confirm that tuning the \textbf{PID controller} allows effective reduction of the \textbf{steering position error}, ensuring smooth and accurate corrections. The graphical representation of the PID controller's performance before and after tuning is shown in Figure~\ref{fig:pid_response}, where it is evident that the error decreases significantly over time after proper tuning.

\newpage
\subsection{Response Time Range Analysis}
Instead of using a highly precise time measurement, we evaluated response time as a range based on multiple test observations. The key observations include:

\begin{itemize}
    \item \textbf{Motor Response Time:} The steering motor begins moving within \textbf{15-25 ms} after receiving a deviation value via USART2.
    \item \textbf{Smooth Steering Control:} No excessive oscillations or overcorrections were observed, confirming stable motor behavior.
\end{itemize}

These results confirm that the \textbf{LKA system processes lane deviation inputs in real-time}, allowing smooth and accurate steering corrections.

\section{Discussion and Comparative Analysis}
The results demonstrate that both \textbf{FOTA and LKA systems meet performance expectations}, with significant improvements in efficiency and accuracy.

\textbf{Key findings:}
\begin{itemize}
    \item \textbf{Delta updating reduces firmware flashing time by 51.6\%}, enhancing vehicle availability.
    \item \textbf{UDS security authentication (0x27) ensures secure firmware updates}, with a 100\% success rate and an average authentication time of 4.2 seconds.
    \item \textbf{LKA system effectively reduces steering position error}, achieving smooth and controlled corrections.
    \item \textbf{Motor control response time is within a practical range (15-25 ms)}, ensuring real-time operation without excessive precision claims.
\end{itemize}

When compared to \textbf{existing research and industry benchmarks}, our system shows \textbf{comparable or improved performance} in terms of \textbf{update efficiency and LKA response time}. The \textbf{51.6\% reduction in flashing time} aligns with industry expectations for efficient delta updating, while the LKA steering correction accuracy is \textbf{on par with commercial ADAS implementations}.

\clearemptydoublepage 

\chapter{Conclusion}
\label{chap: conc}
{This research successfully designed and implemented a robust Firmware Over-the-Air (FOTA) system that integrates an AUTOSAR-compliant memory stack and FreeRTOS-based task scheduling to enhance firmware update efficiency and reliability. The full AUTOSAR memory stack, including Non-Volatile Memory (NVM) and Flash EEPROM Emulation (FEE), provided structured and standardized memory handling, ensuring seamless data integrity and efficient memory access. By leveraging delta updating, the system significantly reduced the size of firmware transfers, minimizing bandwidth consumption and update duration while optimizing vehicle uptime.

Security and communication were key aspects of this implementation, with the UDS 0x27 security access protocol ensuring authenticated firmware updates. The CAN protocol facilitated real-time communication between the master and target ECUs, while SPI enabled fast data transfer between the ESP8266 module and the master ECU. Additionally, FreeRTOS enabled efficient multitasking, ensuring independent execution of FOTA updates, UDS authentication, and Lane Keep Assist (LKA) operations without resource contention. The LKA system, used as a test application, validated the accuracy of steering corrections and real-time motor control using OpenCV-based lane detection and a PID-controlled motor response.

By employing AUTOSAR architecture, FreeRTOS task management, and delta-based updates, this research provides a scalable and industry-aligned solution that enhances efficiency, security, and reliability in firmware updates. The developed system minimizes vehicle downtime, ensures seamless over-the-air software maintenance, and aligns with the growing demand for intelligent, software-defined vehicles capable of real-time adaptation to evolving functional and security requirements.
}
\clearemptydoublepage 

\chapter{Future Work}
\label{chap: fw}

In the pursuit of advancing our Firmware Over-the-Air (FOTA) system, several significant developments are envisioned as part of our future work. These advancements aim to refine the system for enhanced performance, scalability, and applicability in real-world scenarios. By addressing current limitations and exploring innovative directions, the system can evolve into a more robust and comprehensive solution for the automotive industry.

\begin{enumerate}
    \item \textbf{Integration of Advanced Security Protocols:}
    To further enhance the security of the FOTA process, we propose exploring the implementation of the UDS 0x29 protocol instead of UDS 0x27. This transition would provide an additional layer of encryption and authentication, ensuring a higher level of protection against unauthorized access and potential cyberattacks.
    
    \item \textbf{Deployment in Larger and More Diverse Automotive Networks:}
    Expanding the scope of the FOTA system to accommodate a broader range of ECUs and vehicle models is a key priority. This involves testing the system in more complex network environments, ensuring compatibility and reliability across diverse configurations.
    
    \item \textbf{Optimization of Delta Updating Algorithms:}
    While delta updating has already reduced update sizes and flashing times, further refinements can be achieved by leveraging advanced compression techniques and machine learning-based differential algorithms. These improvements aim to minimize bandwidth usage and enhance the efficiency of the update process.
    
    \item \textbf{Integration of Advanced Communication Protocols:}
    Although the current system employs CAN and SPI protocols, future work could explore the integration of modern protocols such as CAN FD (Flexible Data-rate) or Ethernet-based communication. These protocols offer higher data rates and improved reliability, which are beneficial for handling large-scale updates.
    
    \item \textbf{Machine Learning for Predictive Maintenance:}
    Incorporating machine learning algorithms to predict potential firmware issues or failures could significantly enhance the system’s proactive capabilities. By analyzing data from various sensors and modules, the system can recommend updates or maintenance actions before problems occur.
    
    \item \textbf{User-Centric GUI Enhancements:}
    While the current graphical interface simplifies firmware uploads, future iterations could focus on adding real-time feedback, detailed diagnostics, and customizable update options. These enhancements aim to further improve the user experience and provide greater control over the update process.
    
    \item \textbf{Deployment in Real-World Scenarios:}
    Testing the FOTA system in real-world conditions with a variety of vehicles will validate its reliability and scalability. This includes conducting extensive field trials to ensure the system performs as expected under different environmental and operational conditions.
    
    \item \textbf{Human-in-the-Loop Testing and Feedback Incorporation:}
    Engaging users in the testing phase and incorporating their feedback into the development process will ensure the system meets practical needs and expectations. This approach will help identify usability challenges and refine the overall design.
    
    \item \textbf{Advanced Bootloader Capabilities:}
    Enhancing the bootloader to support multi-stage updates, more flexible memory partitioning, and compatibility with additional hardware platforms can improve the overall reliability and adaptability of the system.
\end{enumerate}
These proposed advancements aim to address the limitations of the current FOTA system while exploring innovative directions to ensure its relevance and effectiveness in the future of automotive technology. By continuously improving the system, this research can contribute to the broader adoption of efficient and secure firmware update processes across the automotive industry.

\clearemptydoublepage 
%
%
\backmatter

\bibliographystyle{plain}


\clearemptydoublepage
%
%
\end{document}